\documentclass[12pt,preprint]{aastex}

\shorttitle{Fine spacings and separation ratios of
low-degree $p$ modes}

\shortauthors{Basu et al.}

\begin{document}

\title{Solar abundances and helioseismology: fine structure spacings
and separation ratios of low-degree $p$ modes}

\author{Sarbani Basu}

\affil{Department of Astronomy, Yale University, P.O. Box 208101, New
Haven, CT 06520-8101; sarbani.basu@yale.edu}

\author{William~J.~Chaplin, Yvonne~Elsworth}

\affil{School of Physics and Astronomy, University of Birmingham,
Edgbaston, Birmingham B15 2TT, U.K.; w.j.chaplin@bham.ac.uk,
y.p.elsworth@bham.ac.uk}

\author{Roger~New}

\affil{Faculty of Arts, Computing, Engineering and Sciences, Sheffield
Hallam University, Sheffield S1 1WB, U.K.; r.new@shu.ac.uk}

\author{Aldo~M.~Serenelli}

\affil{Institute for Advanced Study, Einstein Drive, Princeton, NJ
08540; aldos@ias.edu}

\and

\author{Graham~A.~Verner}

\affil{School of Physics and Astronomy, University of Birmingham,
Edgbaston, Birmingham B15 2TT, U.K.; gav@bison.ph.bham.ac.uk}

\begin{abstract}

We have used 4752 days of data collected by the Birmingham
Solar-Oscillations Network (BiSON) to determine very precise
oscillation frequencies of acoustic low-degree modes that probe the
solar core. We compare the fine (small frequency) spacings and
frequency separation ratios formed from these data with those of
different solar models. We find that models constructed with low
metallicity are incompatible with the observations.  The results
provide strong support for lowering the theoretical uncertainties on
the neutrino fluxes. These uncertainties had recently been raised due
to the controversy over the solar abundances.

\end{abstract}


\section{Introduction}
\label{sec:intro}

One of the key inputs used in constructing models of stars, including
the Sun, is the heavy element abundance, $Z$, or the ratio of the
abundance of heavy elements to that of hydrogen $Z/X$. Helioseismic
studies of solar models constructed with the Grevesse \& Sauval (1998;
henceforth GS) abundance, and the even earlier Grevesse \& Noels
(1993) abundances, show a remarkable agreement with the Sun (e.g.,
Christensen-Dalsgaard et al. 1996; Bahcall et al. 1997; Morel et al.
1999; Basu, Pinsonneault \& Bahcall 2000, etc.). The situation has
changed in the recent years with the claim by Asplund et al (2001),
Allende Prieto, Lambert \& Asplund (2001, 2002), Asplund (2004),
Asplund et al. (2004), Asplund et al. (2005), that the oxygen, carbon
and nitrogen abundance in the Sun needs to be reduced. The reduction
in the oxugen abundance implies the reduction of the abundances of
other elements such as Ne, Ar also. This reduces the solar heavy
element abundance from the GS value of $Z/X=0.0229$ to
$Z/X=0.0176$. Asplund, Grevesse \& Sauval (2005; henceforth AGS)
presented a new table of abundances based on the above reduction.  In
addition, based on Asplund (2000) they reduced the silicon abundance
(and thus lowered the abundance of all the meteoritic elements), and
this reduced $Z/X$ further to $0.0166$.  Solar models constructed with
the new, low abundances do not satisfy the helioseismic constraints of
the position of the convection-zone base and the helium abundance in
the convection zone. Models with the lower heavy-element abundances
are found to have too shallow a convection zone (henceforth CZ), and a
helium abundance, $Y$, that is too low. Furthermore, the sound-speed
and density profiles of the models do not match those of the Sun
either (Bahcall \& Pinsonneault 2004; Basu \& Antia 2004; Bahcall et
al.~2005a, 2005b, 2005c, 2006; Delahaye \& Pinsonneault 2006). By
looking at the structures of the main heavy-element ionization zones
in the solar convection zone Antia \& Basu (2006) argue that the solar
heavy element abundance has to be high, i.e., close to the old GS
values. Pinsonneault \& Delahaye (2006) by considering the physics of
stellar atmospheres also conclude that high CNO abundances are
favored.

Almost all the helioseismic tests of solar models constructed with the
new abundances have focused on the solar convection zone. Few studies
have looked at what happens to the solar core.  Where the core has
been considered it has been through results of inversions, where
inversion data on the core have been combined with results pertaining
to the other parts of the Sun to give RMS deviations of the
sound-speed and density profiles of the models with respect to those
of the Sun (e.g., Bahcall et al. 2005a, 2006). In this work, we
concentrate on the solar core.  The solar core is, of course, where
all the nuclear reactions take place, and where all the energy and
neutrinos are released -- hence the interest in what happens to the
core when solar metallicity is reduced.

Only modes of low degree ($l=0$-3) can be used to determine the
structure of the solar core.  In this paper we report a study in which
observed low-degree frequencies have been compared with those of
several standard solar models constructed with different abundances.
The solar $p$-mode data come from observations made by the
ground-based Birmingham Solar-Oscillations Network (BiSON; Chaplin et
al. 1996). The BiSON instruments make disc-averaged observations of
the Sun in Doppler velocity. These observations have provided the
field of helioseismology with high-quality, long-term monitoring of
the low-$l$ modes. Here, we use frequencies determined to excellent
precision from a BiSON timeseries lasting 13 years.

The aim of this paper is to compare and contrast the frequencies of
different solar models with those of the Sun in order to determine
whether we can distinguish between models made with the older, higher
abundances on the one hand and those made with the newer (AGS)
abundances on the other. Our aim is then to see if we can rule out one
set of abundances, thereby reducing the uncertainty in the solar
abundances.  This is particularly relevant for the study of solar
neutrinos. Recently the theoretical uncertainty on solar neutrinos had
been raised because of the uncertainty in the solar abundances
(Bahcall \& Serenelli 2005, Bahcall et al. 2006). The uncertainties in
the values of the heavy-element abundances of the Sun are the largest
source of the theoretical uncertainty in calculating the $p$-$p$,
$pep$, $^8B$, $^{13}N$, $^{15}O$, and $^{17}F$ solar neutrino fluxes
(Bahcall \& Serenelli 2005).  The difference between the GS and AGS
abundances is larger than the error-bars on each of the abundances,
and hence Bahcall \& Serenelli (2005) and Bahcall et al. (2006)
calculated the uncertainties in the neutrino fluxes by designating the
difference between GS and AGS values as the uncertainty in the
composition.  In this work we try to examine if data on the low-degree
modes are of sufficient quality to distinguish between models with the
older, higher abundances and those with the newer, lower abundances
and hence to thereby reduce the neutrino uncertainties to pre-AGS
levels.

A simple comparison of the frequencies of the Sun and solar models is
often not very informative. One reason is that all the modes sample
the outer layers of the Sun, and hence even low-degree modes do not
have localized information on the solar core. Another reason is that
the physics used in the solar models does not reproduce the structure
of the near-surface layers very well: the main culprit is the very
basic treatment of convection, generally using the mixing length
theory, which breaks down close to the surface, another factor is that
the assumption of adiabaticity, used in the calculation of the
frequencies of solar models, breaks down near the solar surface. Both
these errors, and other uncertainties relating to the solar surface,
give rise to frequency errors that are functions of frequency alone
(e.g., Cox \& Kidman 1984; Balmforth 1992 and discussion in
Christensen-Dalsgaard 2002).  To study the core one therefore,
normally compares the so-called \emph{fine} or \emph{small frequency
spacings} of the low-$l$ $p$ modes given by the combination.
 \begin{equation}
 d_{l\,l+2}(n) = \nu_{n,l} - \nu_{n-1,l+2},
 \label{eq:fine}
 \end{equation}
where $\nu$ is the frequency of a mode of degree $l$ and radial order
$n$.  The frequencies $\nu_{n,l}$ and $\nu_{n-1,l+2}$ are very similar
and hence are affected in a similar way by near-surface effects. By
taking this difference in frequency a large part of the effects from
the near-surface uncertainties thus cancels out.

The fine spacings are determined predominantly by the sound-speed
gradient in the core. Using the asymptotic theory of $p$-modes it can
be shown that (see e.g., Christensen-Dalsgaard \& Berthomieu 1991)
\begin{equation}
d_{l\,l+2}(n)\simeq -(4l+6){\Delta_l(n)\over{4\pi^2\nu_{n,l}}}\int_0^R
{dc\over dr}{dr\over r},
\label{eq:smallsep}
\end{equation}
where, $R$ is the solar radius, and  $\Delta_l(n)$ is the large frequency spacing
given by
 \begin{equation}
 \Delta_l(n) = \nu_{n,l} - \nu_{n-1,l},
\label{eq:largesep}
 \end{equation}
which depends inversely on the sound-travel time between the centre
and the surface of the Sun. Given that the gradient of the sound-speed
is large in the core, and $r$ there is small, the integral in
Eq.~\ref{eq:smallsep} is dominated by conditions in the core, and
hence the fine spacings are a useful tool to study conditions in the
solar core.

Although the difference between the frequencies of two modes that have
very similar frequencies does reduce the effect of near-surface
uncertainties, there are some residual effects.  One way of reducing
the effects of the near-surface errors is to use the so-called
\emph{frequency separation ratios}.  The frequency separation ratios
(Roxburgh \& Vorontsov 2003; Ot\'i Floranes, Christensen-Dalsgaard \&
Thompson 2005; Roxburgh 2005) are formed from the fine (small)
frequency spacings and large frequency spacings of the modes.  The
ratios are constructed according to:
 \begin{equation}
 r_{02}(n) = \frac{d_{02}(n)}{\Delta_1(n)},
 ~~~~~~~~r_{13}(n) = \frac{d_{13}(n)}{\Delta_0(n+1)}.
 \label{eq:rats}
 \end{equation}
Since both the small and large spacings are affected in the same manner
by near-surface effects, these ratios are somewhat independent of the
structure of the surface.

In this paper we use data from BiSON to form observational estimates
of the fine spacings, $d_{02}(n)$ and $d_{13}(n)$, and the separation
ratios, $r_{02}(n)$ and $r_{13}(n)$. The observed spacings and ratios
have then been compared with estimates formed from a number of
different solar models. BiSON data on the fine spacings have been used
in several previous studies (e.g., Elsworth et al. 1990; Chaplin et
al. 1997), but those here come from much longer observations and have
far superior precision.

The rest of the paper is organized as follows: the observed data are
described in \S~\ref{sec:data}, the solar models are described in
\S~\ref{sec:models}, we present and explain our results in
\S~\ref{sec:res}, and we discuss the implications of these results to
solar neutrino predictions in \S~\ref{sec:neut}, and we present our
conclusions in \S~\ref{sec:conclu}.

\section{Data}
\label{sec:data}

\subsection{Observations}
\label{sec:obs}

We have used Doppler velocity observations made by the BiSON over the
4752-d period beginning 1992 December 31, and ending 2006 January
3. Mode peaks in the power spectrum of the complete time series were
fitted in the usual manner (e.g., see Chaplin et al. 1999) to yield
estimates of the low-$l$ frequencies.

\subsubsection{Impact of the solar cycle on the Sun-as-a-star frequencies}
\label{sec:cyc}

The BiSON observations are examples of the so-called
\emph{Sun-as-a-star} data: averages over the visible disc of the
perturbations -- here, in Doppler velocity -- associated with the
modes. Chaplin et al. (2005) showed that fine spacings and separation
ratios made from the Sun-as-a-star data are sensitive to the changing
surface activity along the solar cycle. In short, the acoustic
asphericity leaves its imprint on the azimuthally dependent
Sun-as-a-star frequencies.

Some of the mode components are effectively missing from the data
because the sensitivity of the observations to them is so low. This is
a consequence of the visibility of any given $m$ being a strong
function of the angle of inclination, $i$, offered by the star. Extant
Sun-as-a-star observations, such as those of the BiSON, are made from
$i \approx 90$ degrees. This means that only components with $l+m$
even have non-negligible visibility. As such, it is not possible to
estimate directly the frequency centroid; and estimates of the
`frequency' of a mode will be influenced strongly by the 
$|m|=l$
modes, which are most prominent in the data. The Sun-as-a-star fine
spacings are therefore formed from modes with different combinations
of $l$ and $m$, which may suffer different sized frequency shifts
through the solar cycle.

For the $d_{02}(n)$ data, estimates of the $l=2$ frequencies are
dominated by the $|m|=2$ modes. They show significantly larger cycle
shifts than the lone components of the nearby $l=0$ modes (e.g., see
Chaplin et al. 2004a; Jim\'enez-Reyes et al. 2004). The $d_{02}(n)$
therefore decrease in size as levels of surface activity
increase. Differences between the $|m|=3$ and 1 components of,
respectively, the $l=3$ and 1 modes, are less pronounced, and so cycle
shifts arising in the $d_{13}(n)$ are smaller, at fixed frequency,
than in their $d_{02}(n)$ counterparts.

The separation ratios show a similar behavior. For these data,
changes in the large spacing, $\Delta_l(n)$, which lies in the
denominator of Equation~\ref{eq:rats}, are negligible. This is because
$\Delta_l(n)$ is formed from the difference of two frequency estimates
in modes with the same combination of $l$ and $m$. While it is true
that there is a dependence of the mode shifts on frequency, the
separation of the overtones used to form the $\Delta_l(n)$ is
sufficiently small that the impact of this dependence is
modest. Changes to the separation ratios are therefore dominated by
changes to the fine spacings. This means that fractional changes in
the $r_{02}(n)$ are, to first order, the same as those in the
$d_{02}(n)$ [and likewise for the $r_{13}(n)$ and $d_{13}(n)$].

\subsubsection{Correction of the Sun-as-a-star frequencies for the solar cycle}
\label{sec:corr}

Prior to calculation of the fine spacings and separation ratios, we
therefore removed the solar-cycle shifts from the raw fitted low-$l$
frequencies. Our procedure rests on the assumption that variations in
certain global solar activity indices can be used as a proxy for the
low-$l$ frequency shifts, $d\nu_{n\ell}(t)$. We assume the correction
can be parameterized as a linear function of the chosen activity
measure, $A(t)$. When the 10.7-cm radio flux (Tapping \& De~Tracey
1990) is chosen as the proxy, this assumption is found to be robust
(e.g., Chaplin et al. 2004a) at the level of precision of the data.

Consider then the set of measured eigenfrequencies, $\nu_{n\ell}(t)$,
that we wish to correct, extracted from data collected over the
$t=4752\,\rm d$ epoch when the mean level of the 10.7-cm radio flux
was $\left< A(t) \right> = 121 \times 10^{-22}\,\rm
W\,m^{-2}\,Hz^{-1}$. We make the correction to the canonical quiet-Sun
level of the radio flux, $A_{\rm quiet}$, which, from historical
observations of the index, is usually fixed at $64 \times
10^{-22}\,\rm W\,m^{-2}\,Hz^{-1}$ (see Tapping \& DeTracey 1990).  The
magnitude of the solar-cycle correction -- which must be subtracted
from the raw frequencies -- will then be:
 \begin{equation}
 \delta \nu_{n\ell}(t) = g_{\ell} \cdot {\cal F}[\nu] 
                   \cdot [\left< A(t) \right> - A_{\rm quiet}].
 \label{eq:corr}
 \end{equation}
The $g_{l}$ are $l$-dependent factors that calibrate the size of the
shift: recall from Section~\ref{sec:cyc} above that the Sun-as-a-star
shifts alter significantly with $l$.  
Owing to the nature of the Sun-as-a-star data, the change with $l$ 
comes from the spatial dependence of the surface activity.
To determine the $g_l$, we
divided the 4752-day timeseries into 44 independent 108-day
segments. The resulting ensemble was then analyzed, in the manner
described by Chaplin et al. (2004a), to uncover the dependence of the
solar-cycle frequency shifts on the 10.7-cm radio flux. The ${\cal
F}[\nu]$ in Equation~\ref{eq:corr} is a function that allows for the
dependence of the shift on mode frequency. Here, we used the
determination of ${\cal F}[\nu]$ to be found in  Chaplin et
al. (2004a, b).

Uncertainty in the correction is dominated by the errors on the
$g_l$. These errors must be propagated, together with the formal
uncertainties from the mode fitting procedure, to give uncertainties
on the corrected frequencies, $\nu_{n\ell}(t)-\delta
\nu_{n\ell}(t)$. The corrected uncertainties are, on average, about
10\,\% larger than those in the raw, fitted frequencies.

After application of the correction procedure, fine spacings and
separation ratios were calculated according to Equations~\ref{eq:fine}
and~\ref{eq:rats} respectively. Uncertainties on the corrected
frequencies were propagated accordingly to give those in the spacings
and ratios.

\begin{figure*}

 \epsscale{1.0} \plottwo{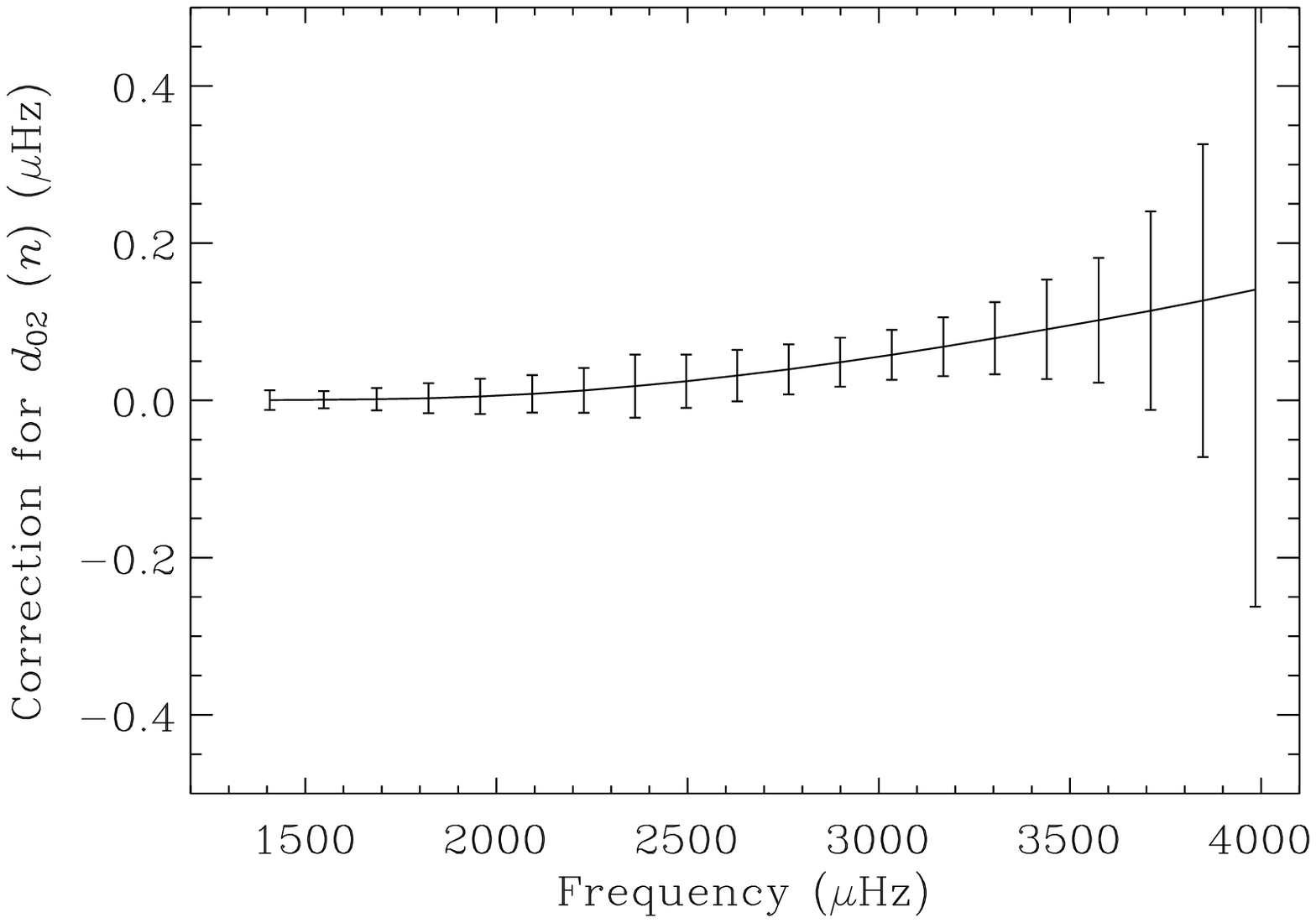}{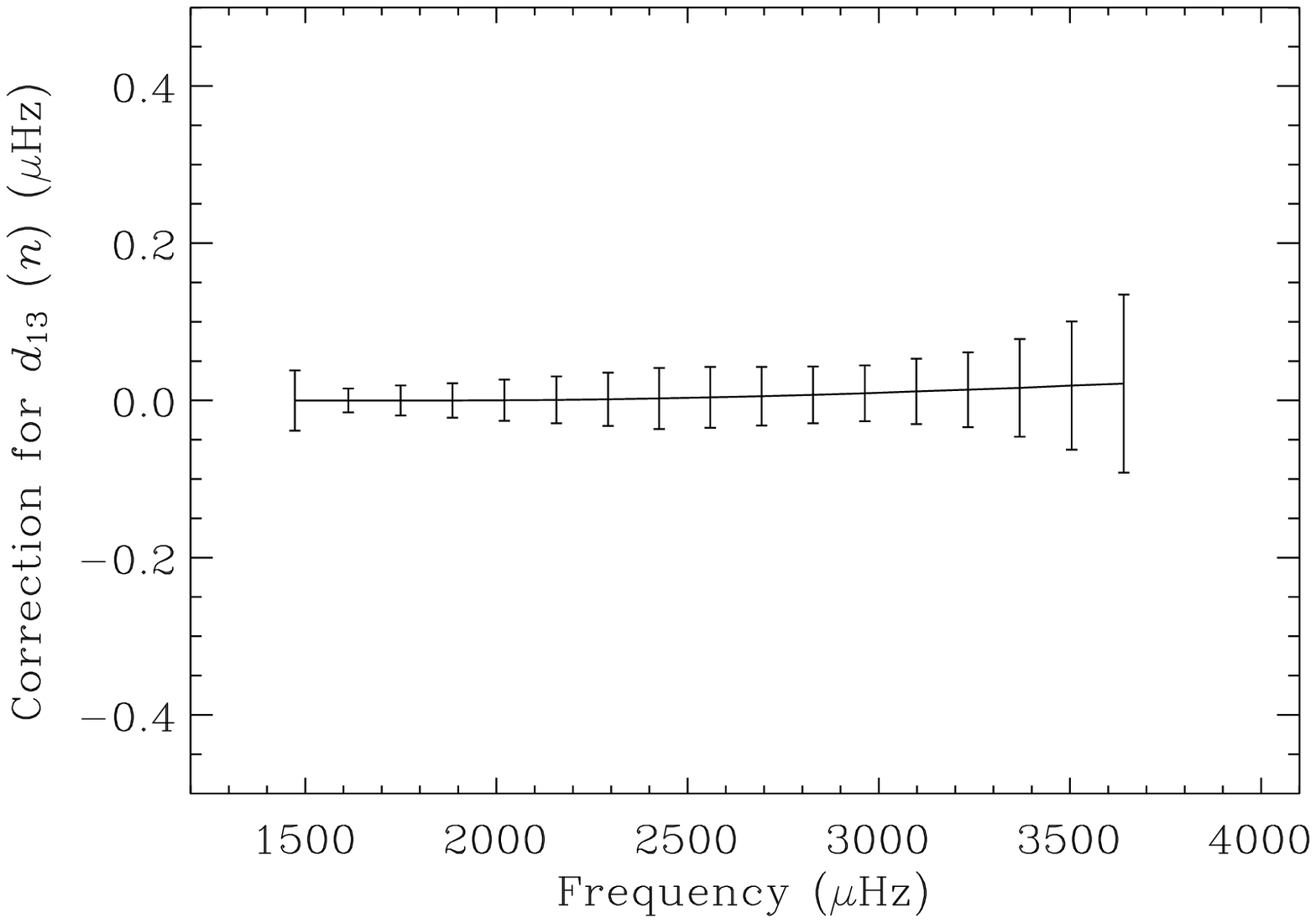}
 
 \caption{Differences between fine spacings made with and without the
solar-cycle correction (in the sense corrected spacings minus raw
spacings). Left-hand panel: difference/correction in
$d_{02}(n)$. Right-hand panel: difference/correction in $d_{13}(n)$.}

 \label{fig:dcorr}

 \end{figure*}

Fig.~\ref{fig:dcorr} shows the differences between fine spacings made
with and without the correction (in the sense corrected spacings minus
raw spacings). The ratio data give similar-looking plots. As indicated
previously, it is the $d_{02}(n)$ (and by implication the $r_{02}(n)$)
that are most affected by the correction procedure.

\subsection{Models}
\label{sec:models}

For the major part of the work we used eight solar models that have
been constructed with different physical inputs. The models are:

\noindent JCDS --- Model S of Christensen-Dalsgaard et
al.~(1996). This model was constructed with surface $Z/X=0.0245$
(Grevesse \& Noels 1993), OPAL(1992) opacities (Rogers \& Iglesias
1992) and the OPAL(1996) equation of state (Rogers, Swenson \&
Iglesias 1996). This model was used because many helioseismological
results in literature are based on this reference model.

\noindent SAC --- Model Seismic$_1$ of Couvidat et al. (2003). It used
OPAL(1996) opacity tables (Iglesias and Rogers 1996), and the
OPAL(1996) equation of state (Rogers, Swenson \& Iglesias 1996). The
model has $(Z/X)=0.02628$.

\noindent BP04 -- Model BP04(Garching) of Bahcall et al. (2005c).
This model was constructed with GS abundances ($Z/X=0.0229$),
OPAL(1995) opacities (Iglesias \& Rogers 1996) and the OPAL(2001)
equation of state (Rogers 2001, Rogers \& Nayfonov 2002).

\noindent BP04+ --- Model BP04+ of Bahcall et al. (2005a). This model
was constructed with abundances from Asplund et al. (2000, 2004),
Asplund (2000), Allende Prieto et al.  ( 2001, 2002). These abundances
imply $Z/X=0.0176$ as opposed to the $Z/X=0.0229$ of GS. Only
abundances of C, N, O, Ne and Ar are lowered in comparison to model
BP04. The rest of the input physics is the same as BP04.

\noindent BS05(OP) --- Model BS05(OP) of Bahcall et al.~(2005c). This
model is similar to model BP04, but has been constructed with
opacities from the OP project (Badnell et al. 2005) instead of the
OPAL opacities.  There are some other subtle differences too, such as
in the treatment of diffusion where, unlike in BP04, each metal has a
different diffusion velocity as derived from Thoul et
al. (1993). Also, unlike in BP04, $^{17}O$ is not burned at all.  The
model was constructed with GS abundances and has a surface
$Z/X=0.02292$.

\noindent BS05(AGS, OP) --- Model BS05(AGS, OP) of Bahcall et
al.~(2005c). This model is similar to model BS05(OP), and was
constructed with AGS abundances, having surface $Z/X=0.01655$.

\noindent BBS05-3 --- Model 3 of Bahcall et al.~(2005b). This model
has the same physics as BS05(OP) and BS05(AGS, OP), but the Ne, Ar,
CNO and Si abundances have been enhanced with respect to the AGS
values. The model has $Z/X=0.02069$.

\noindent S06+(AGS, OP) --- A model constructed specifically for this
work.  It is similar to model BS05(AGS, OP), however, the opacities
have been artificially increased by 13\% in the temperature range of 2
to 5 million Kelvin to get the helioseismically determined position of
the convection-zone base.

We also used three other models to test the effect of heavy elements,
keeping all other input physics the same. These models were
constructed using YREC, the Yale Rotating Evolution Code, in its
non-rotating configuration (Guenther et al. 1992).  All models have
the same physics inputs except the heavy-element abundances.  The
models use the OPAL2001 equation of state, and OPAL(1996) opacities.
The models are:

\noindent YREC(AGS) --- model with surface $Z/X=0.0165$, the AGS value
of surface metallicity.

\noindent YREC(GS) --- model with surface $Z/X=0.0229$, the GS value
of surface metallicity.

\noindent YREC(0.03) --- model with a high metallicity, $Z/X=0.03$,
constructed to check if the effect of metallicity is monotonic.

All models except S06+(AGS, OP) and the three YREC models are published 
models. The models have been calibrated to slightly different values of 
luminosity and radius.  The SAC model has a radius of $6.95936\times 
10^{10}$ cm. All others have a radius of $6.9598\times 10^{10}$ cm.
Model JCDS has a luminosity of $3.8456\times 10^{33}$ ergs/s; 
models BP04, BP04+, BS05(OP), BS05(AGS, OP), BBS05-3, and S06+(AGS, OP) 
have luminosities of $3.8418\times 10^{33}$ ergs/s; and the three YREC 
models have a luminosity of $3.851\times 10^{33}$ ergs/s. The luminosity that the 
SAC model was calibrated to has not been published.

The frequencies for each of these models was calculated by solving
the full set of equations describing stellar oscillations.
The numerical precision of the frequencies
was increased using the Richardson extrapolation technique. Once the frequencies
were calculated, The large and small separations, and the
separation ratios were calculated  using equations \ref{eq:fine}, \ref{eq:largesep}, and
\ref{eq:rats}.  

\section{Results and Discussion}
\label{sec:res}

\begin{figure*}

 \epsscale{1.0}
 \plottwo{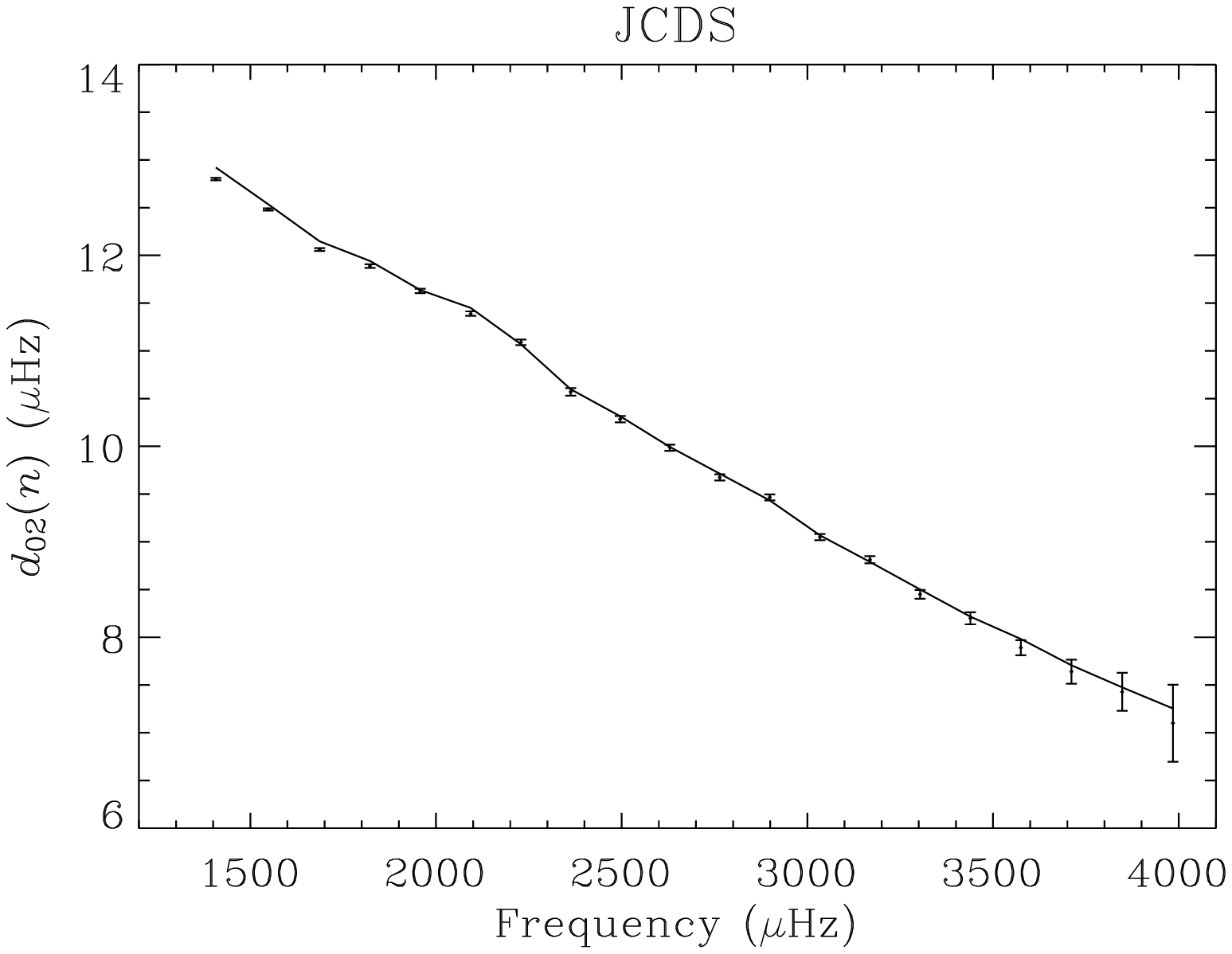}{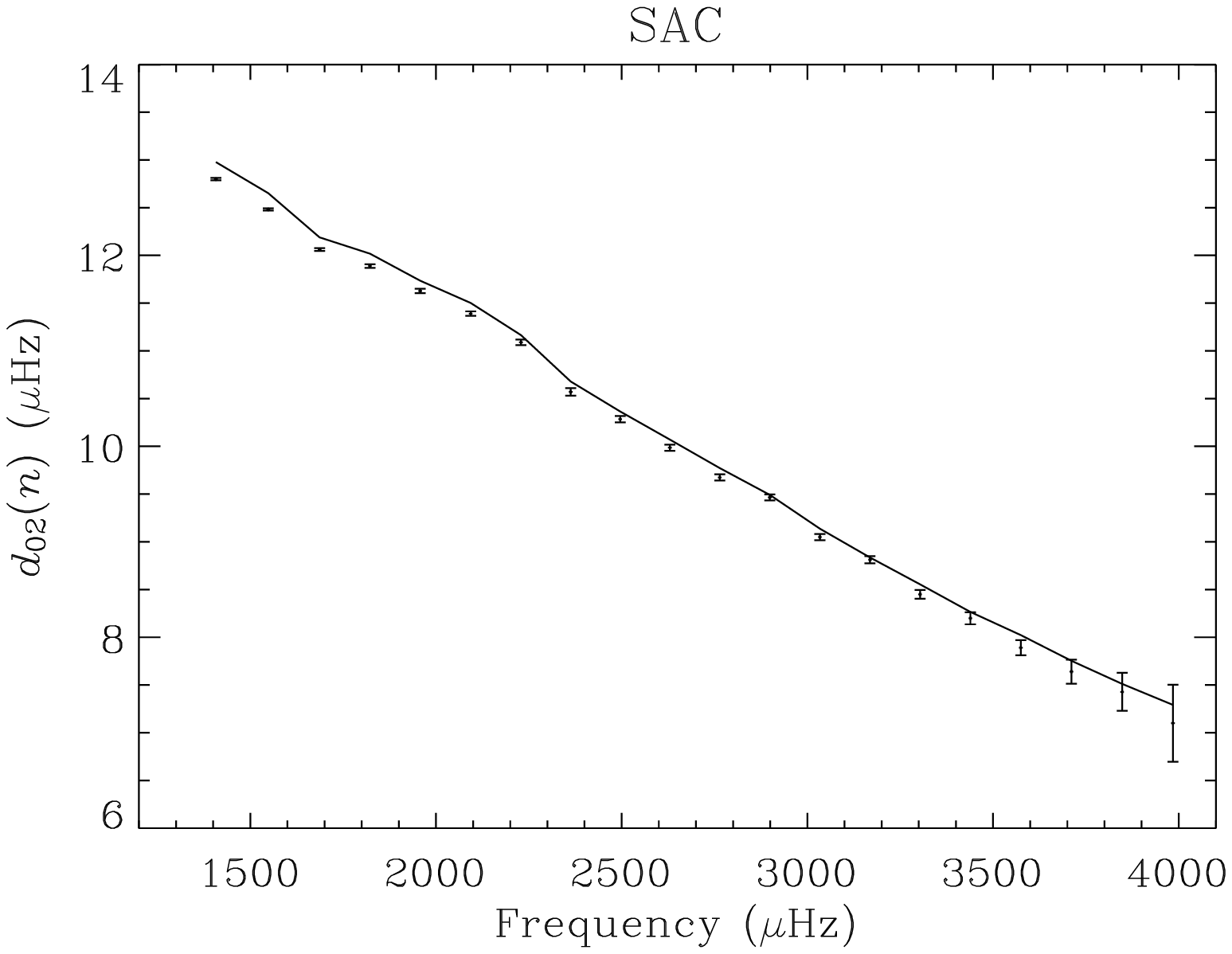}\\
 \plottwo{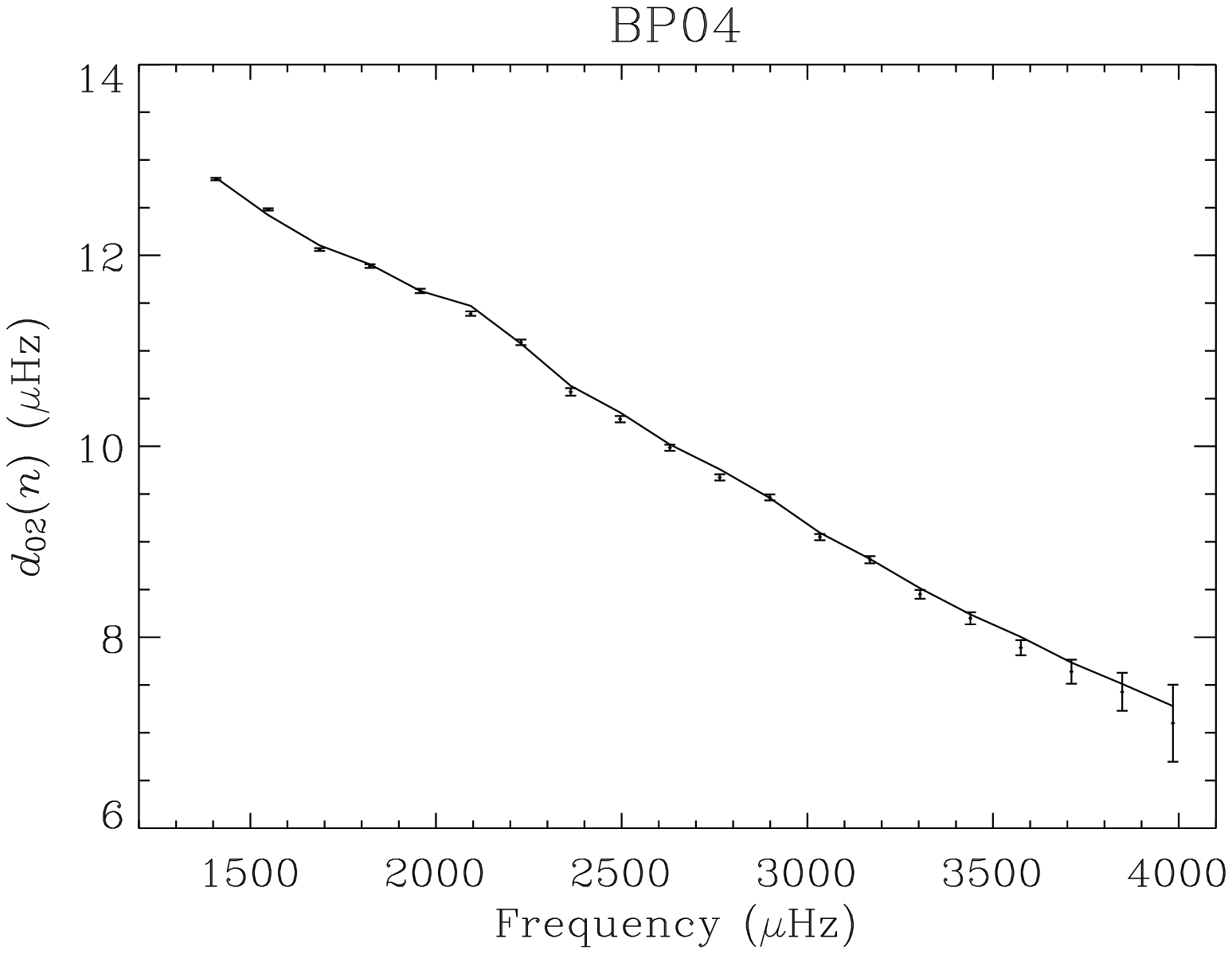}{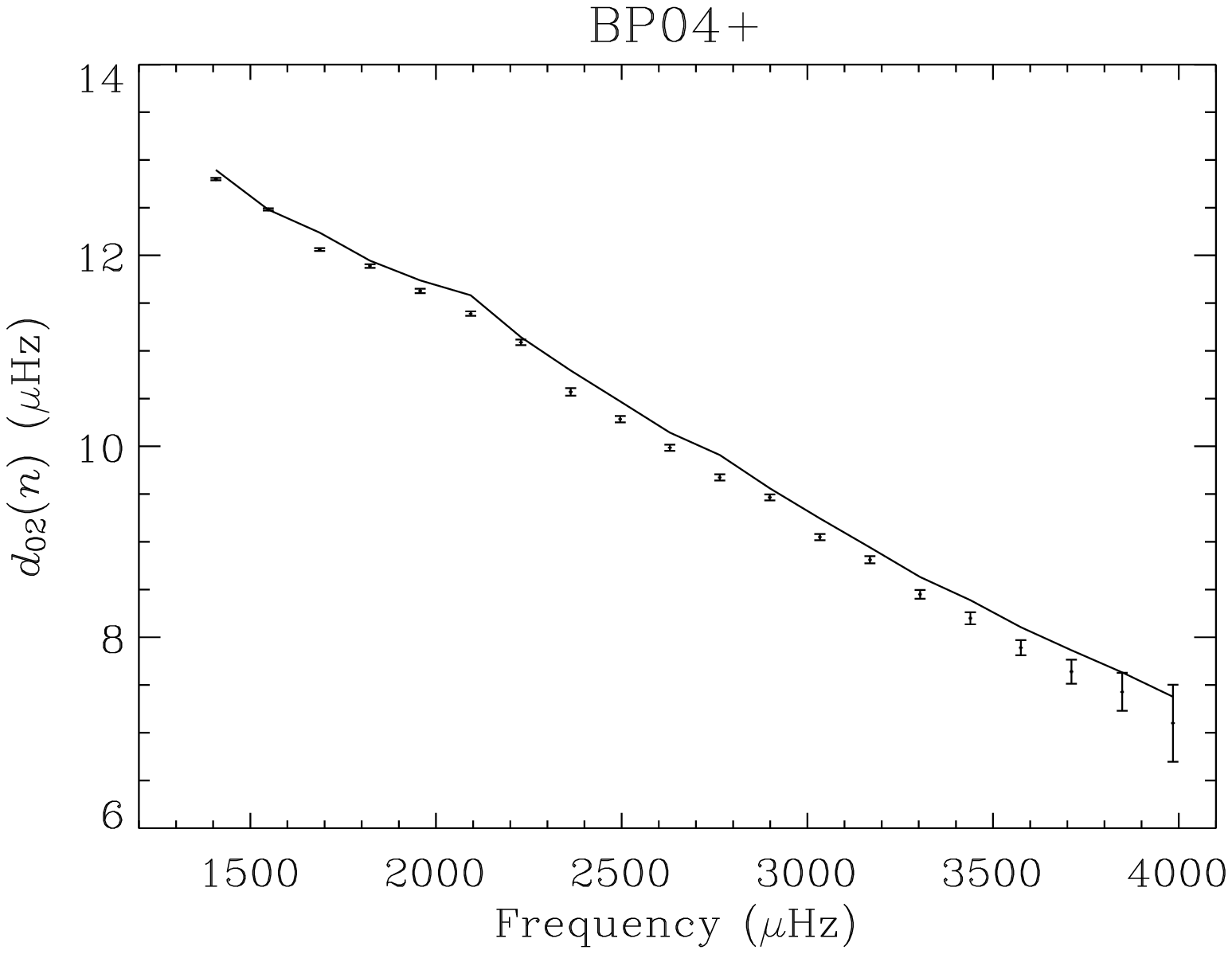}\\
 \plottwo{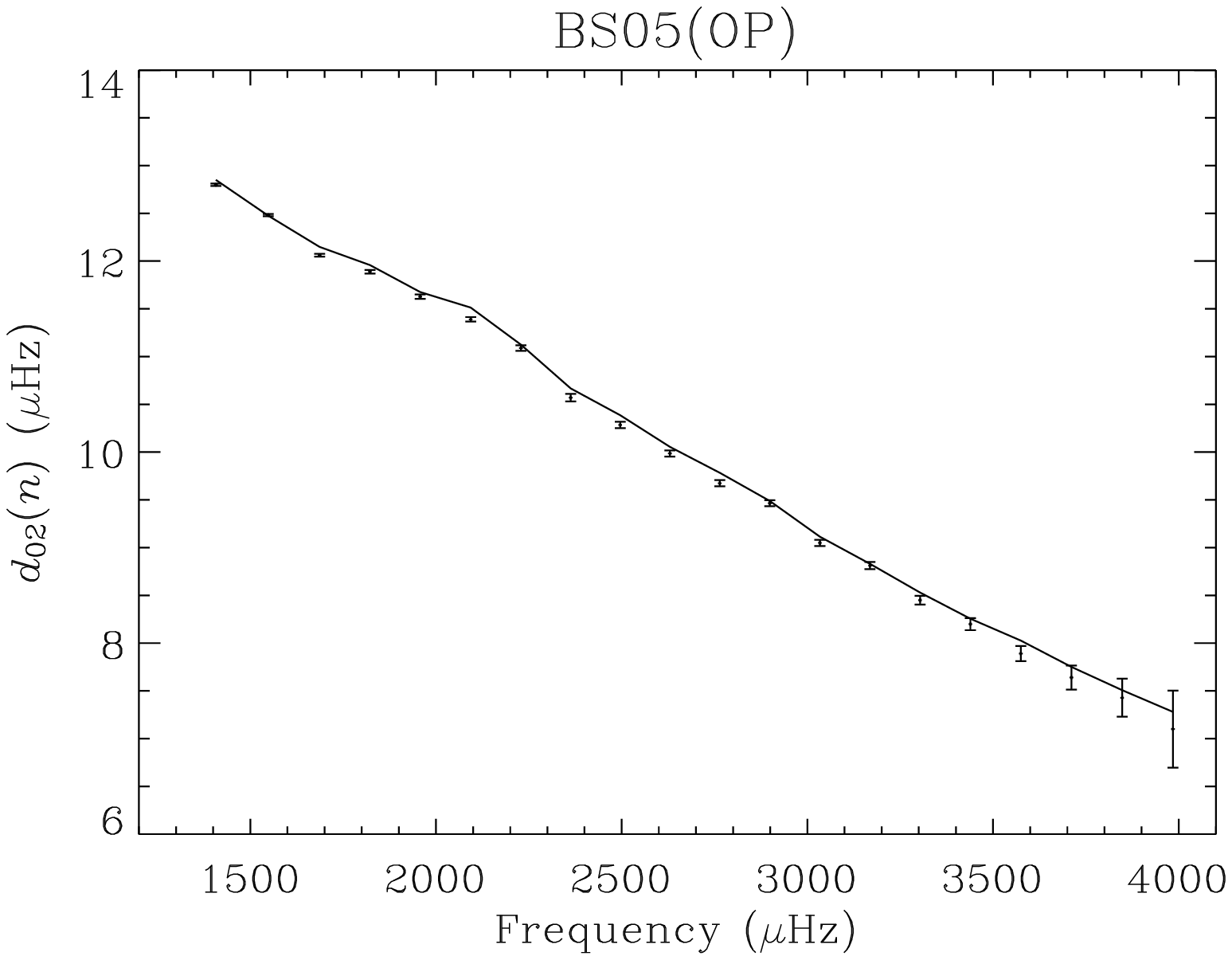}{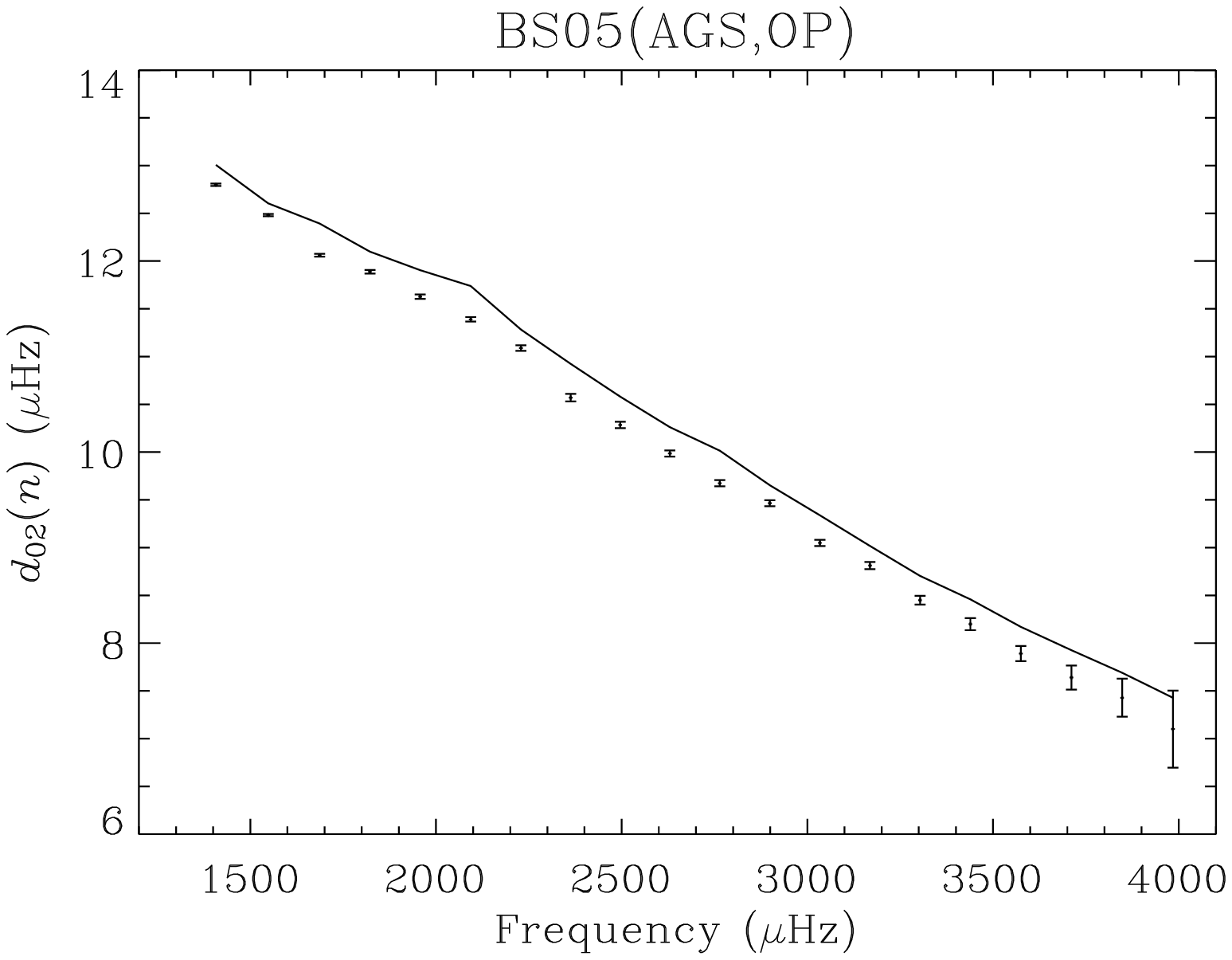}\\
 \plottwo{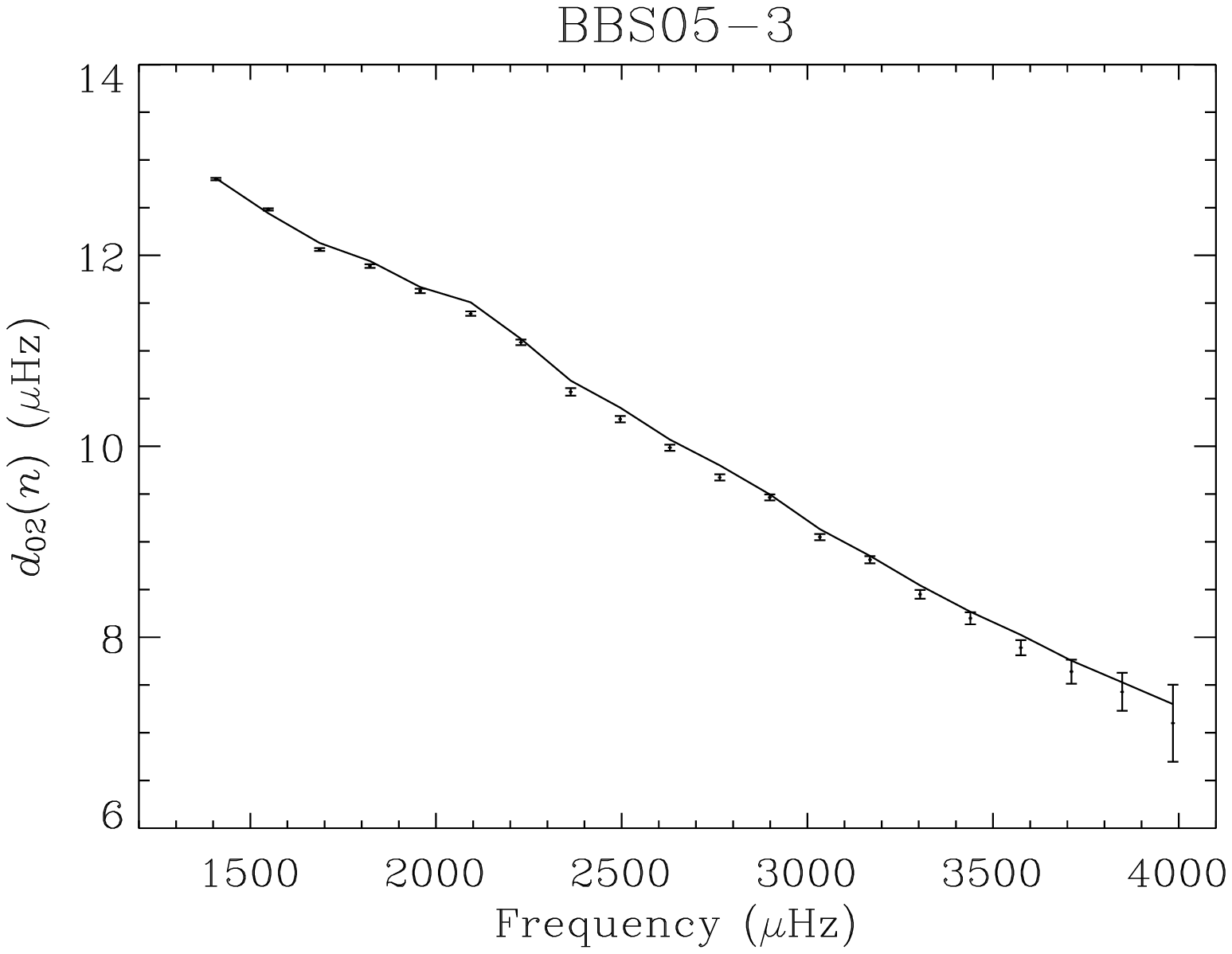}{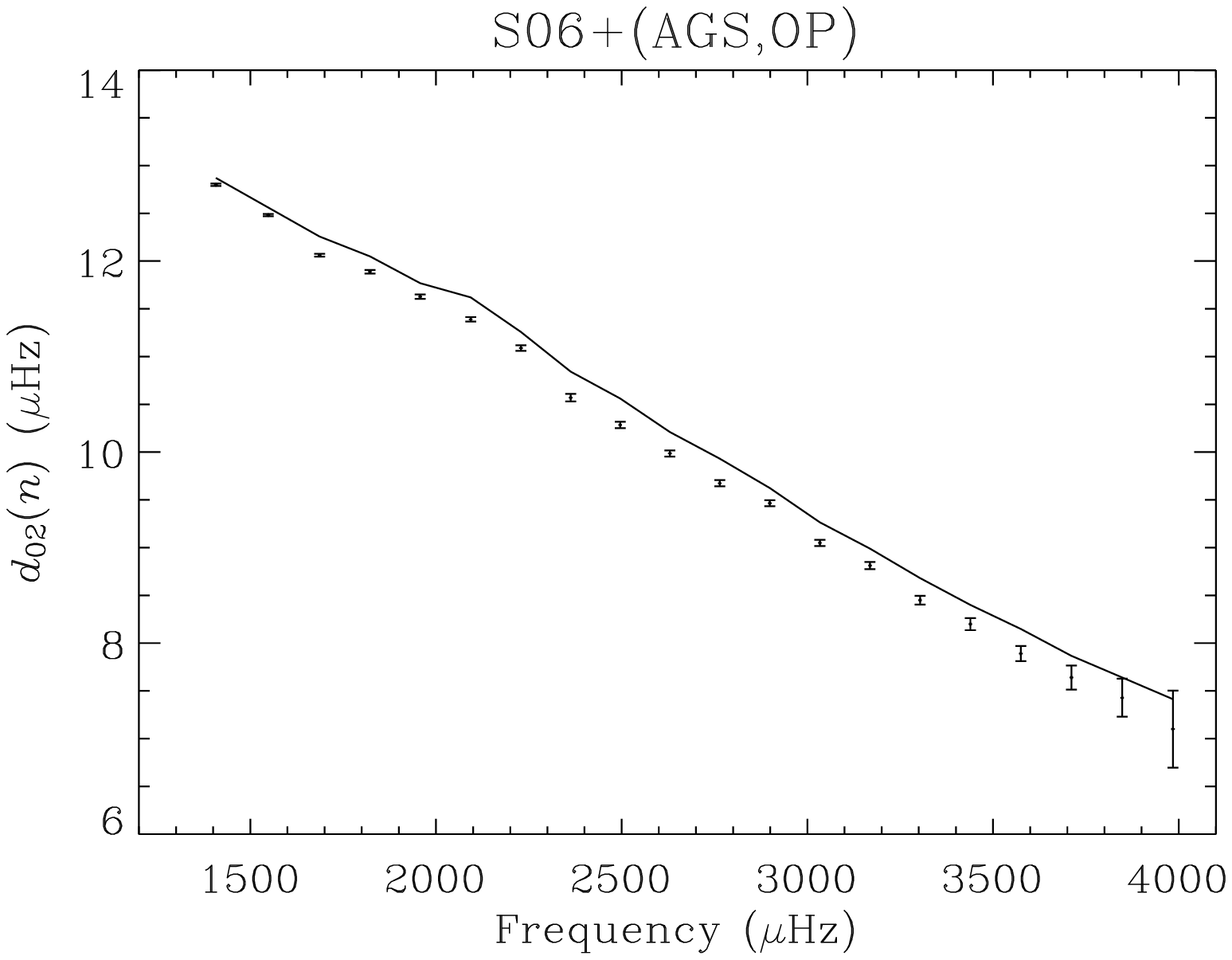}

 \caption{Solid line in each panel: fine spacings $d_{02}(n)$ of solar
 model identified in plot title. Points with error bars: $d_{02}(n)$
 from the corrected BiSON frequencies.}

 \label{fig:d0}

 \end{figure*}

\begin{figure*}

 \epsscale{1.0}
 \plottwo{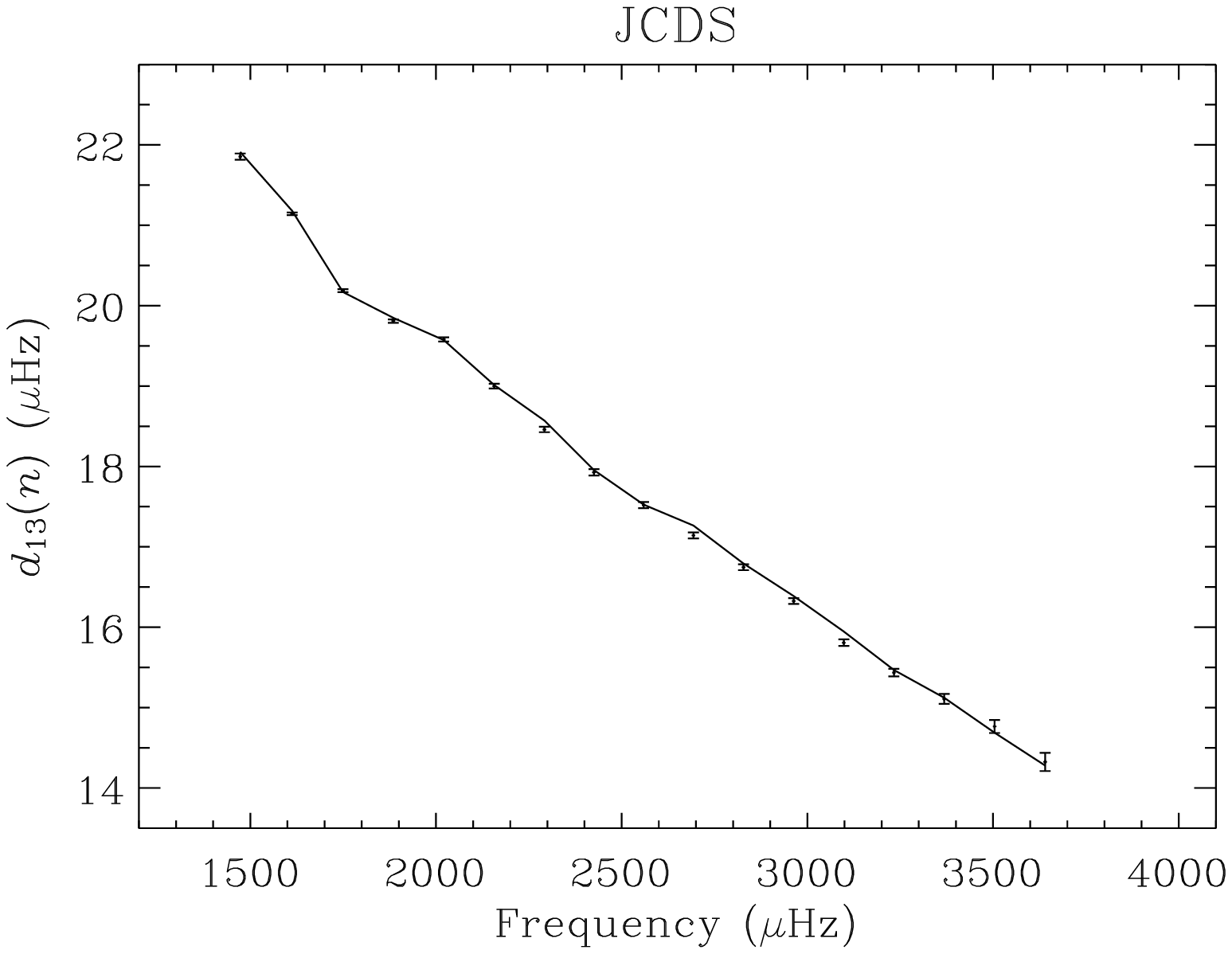}{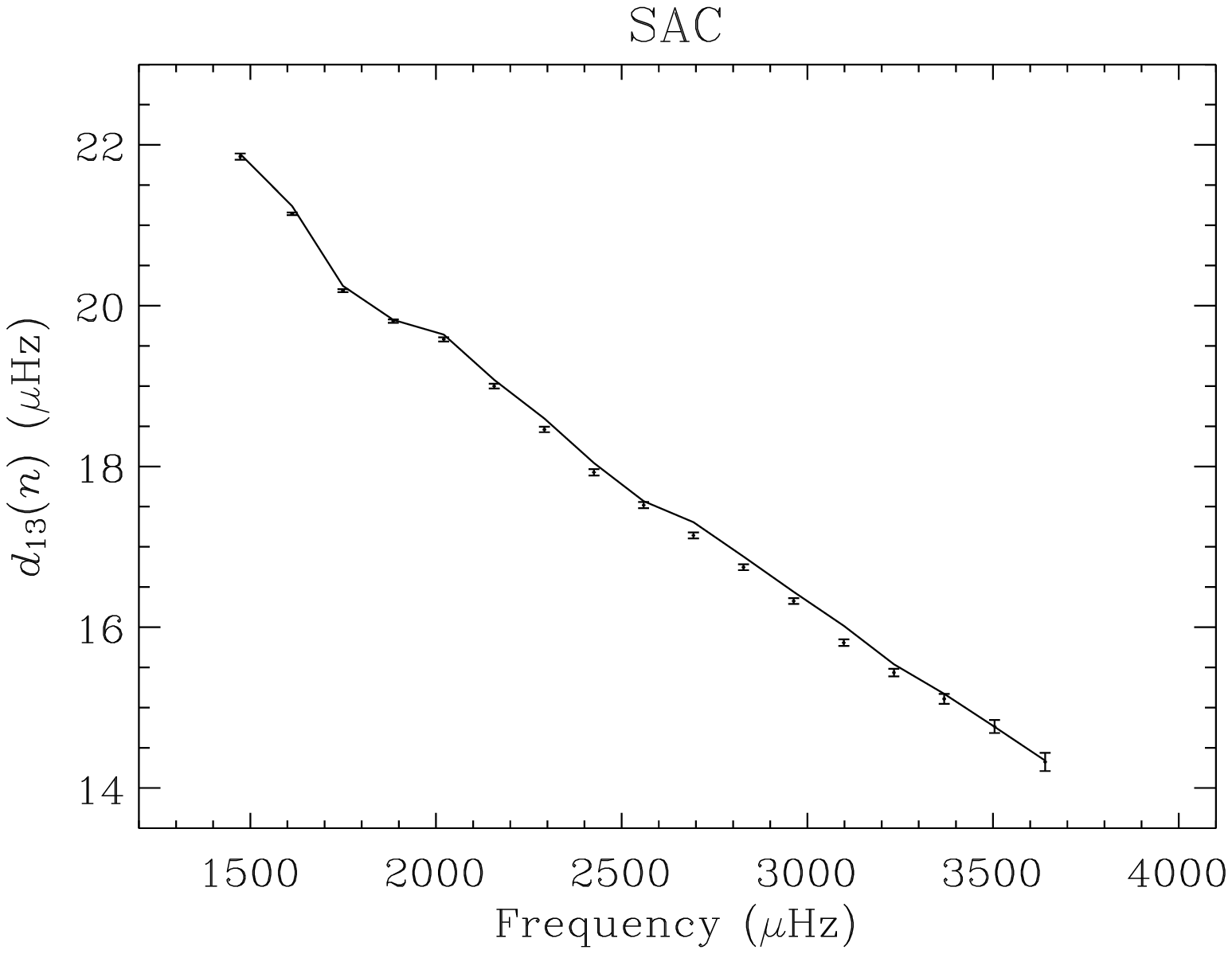}\\
 \plottwo{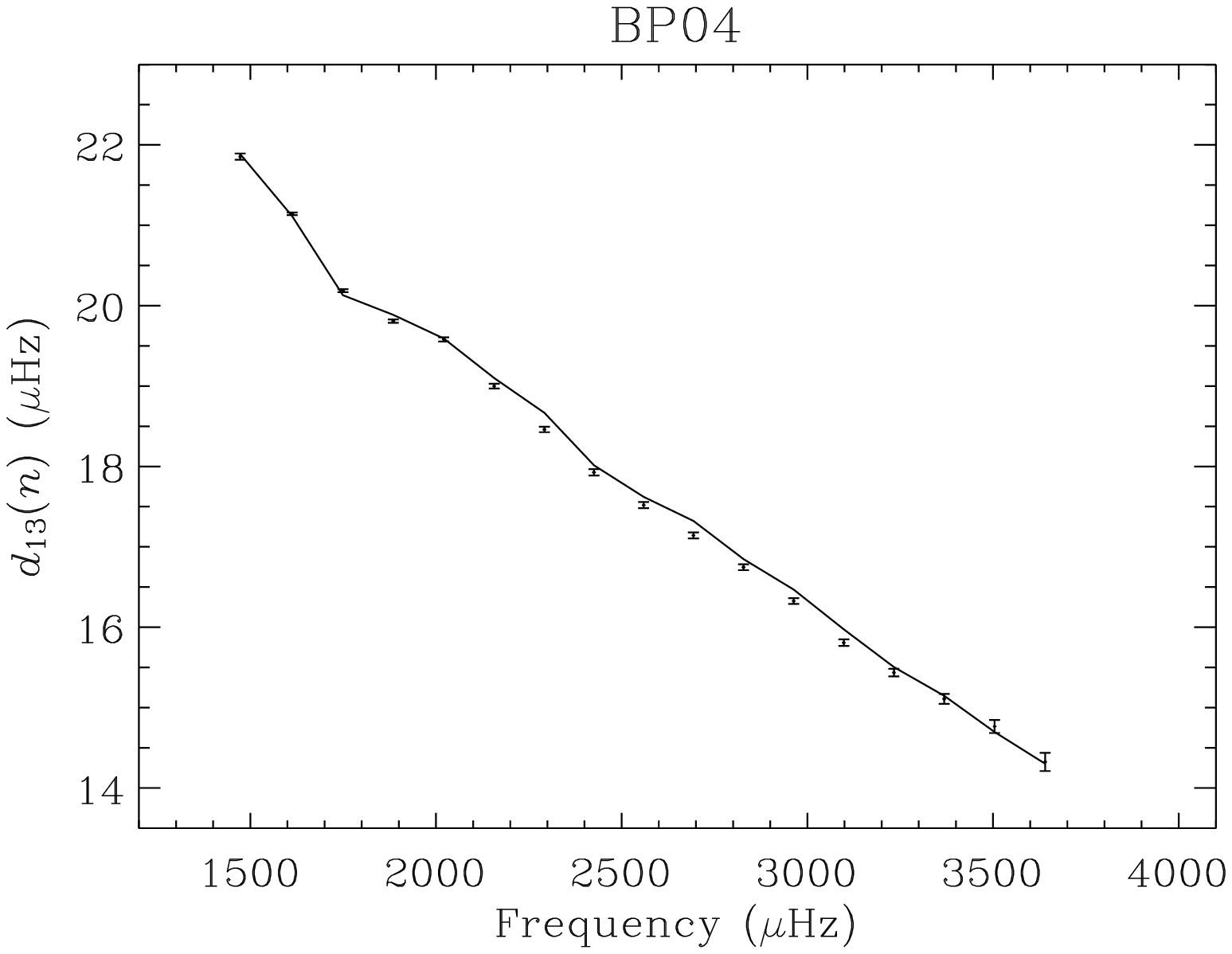}{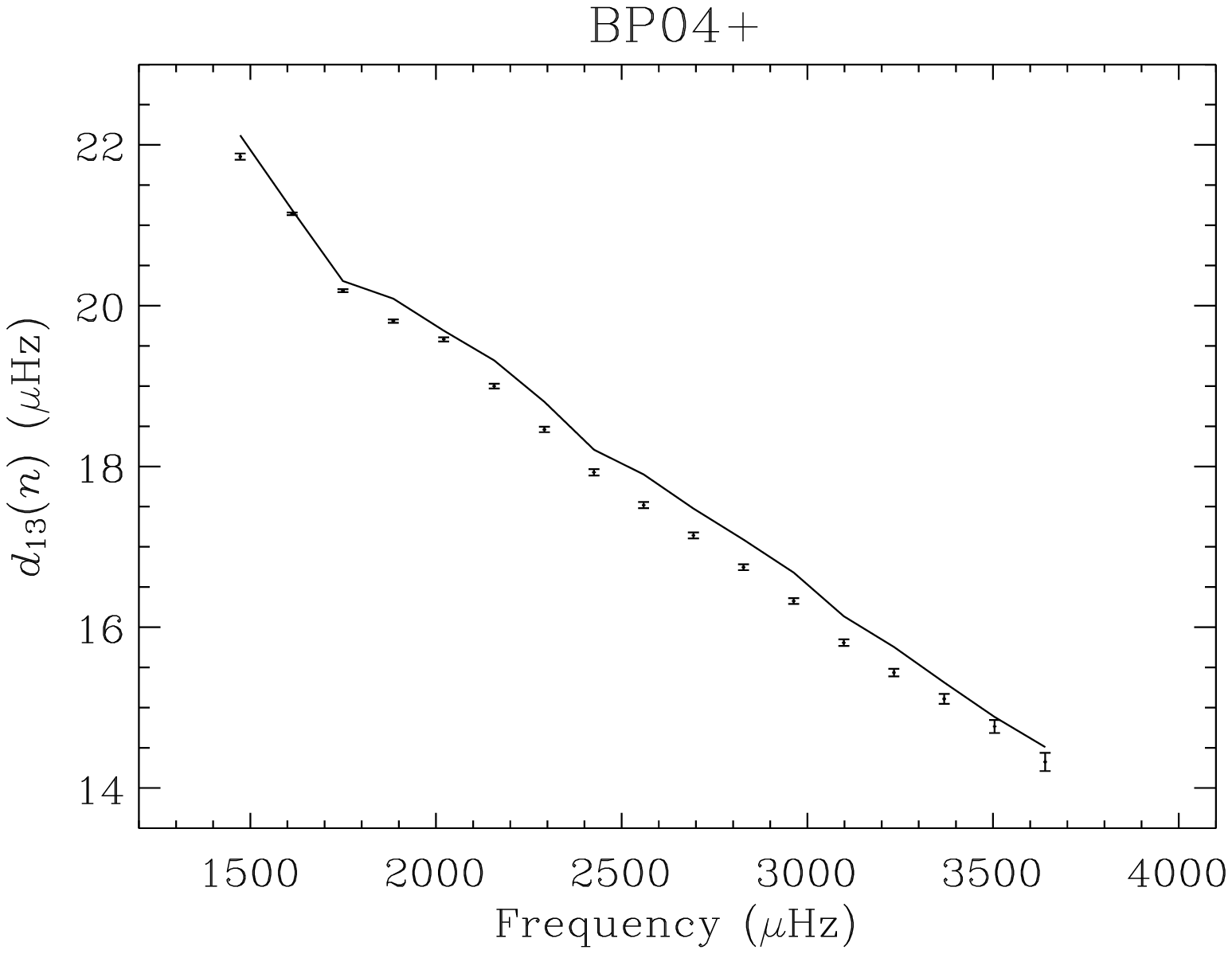}\\
 \plottwo{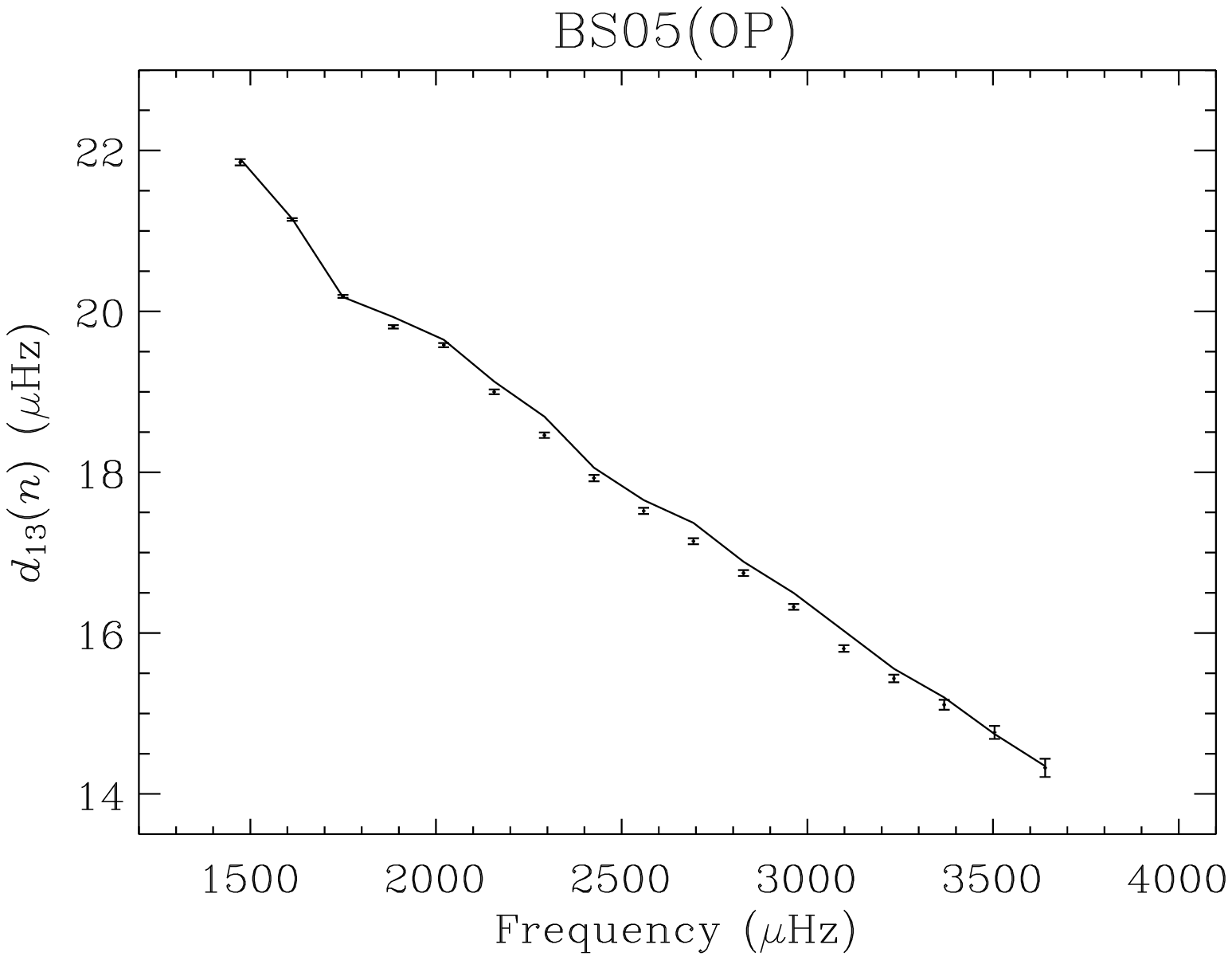}{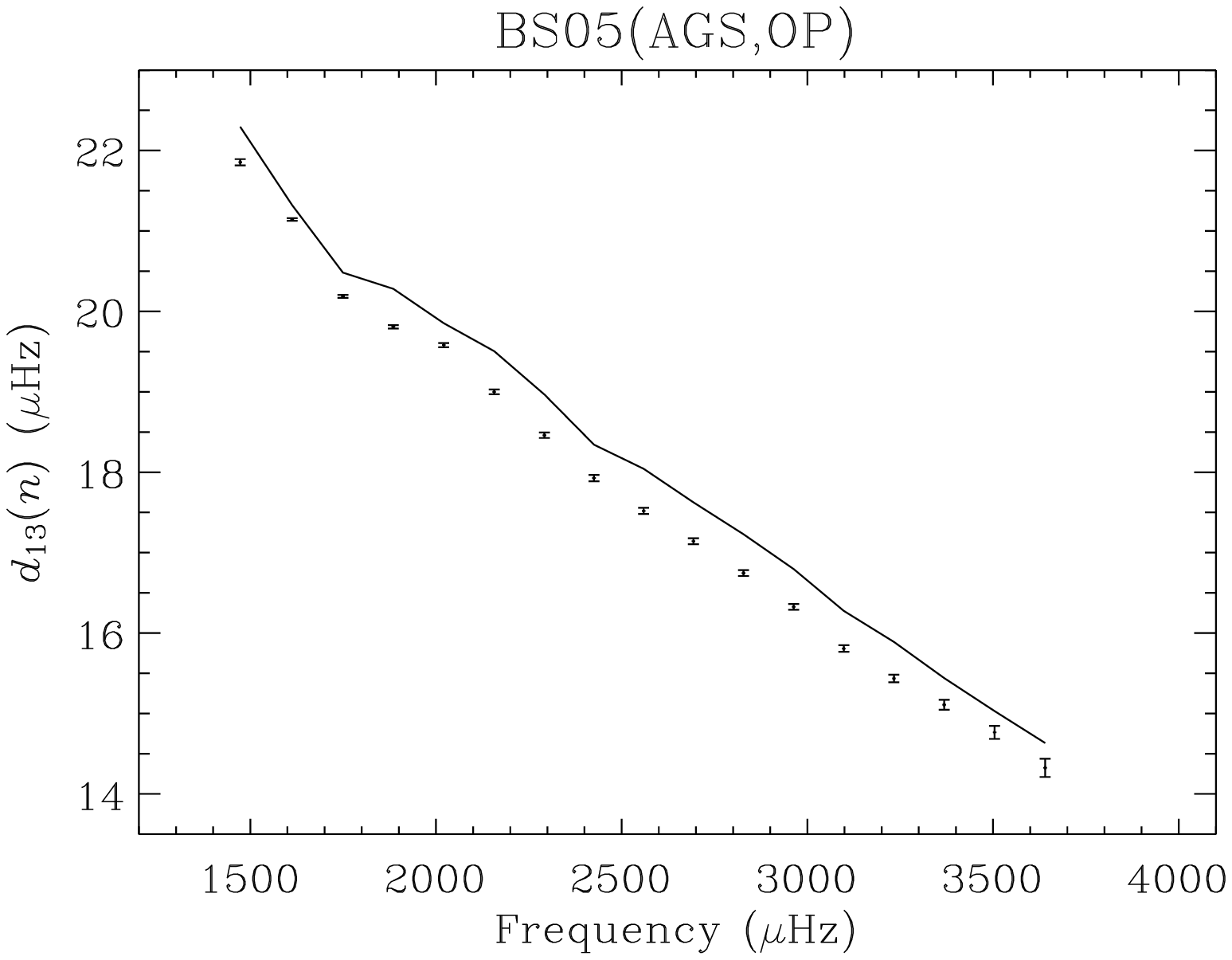}\\
 \plottwo{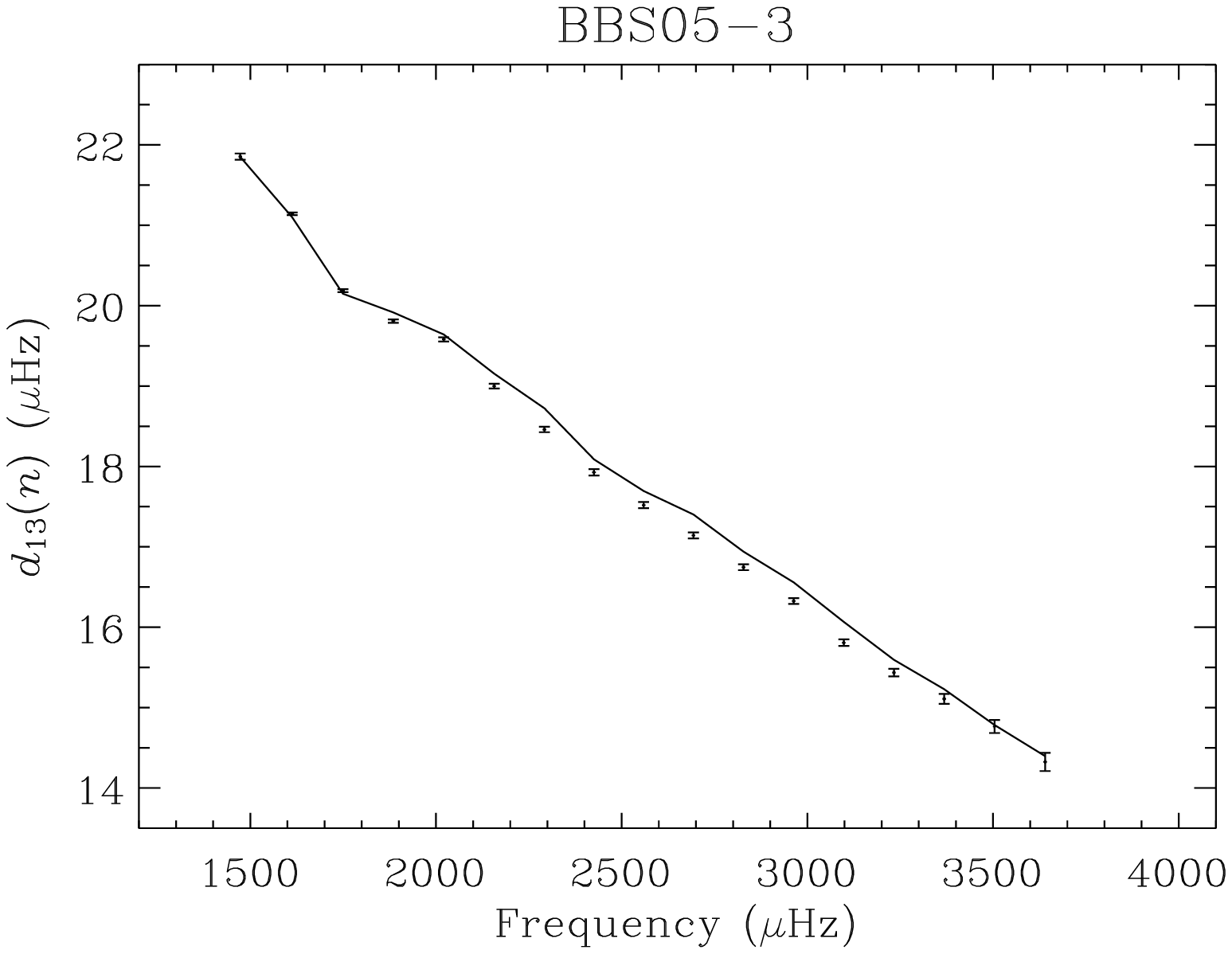}{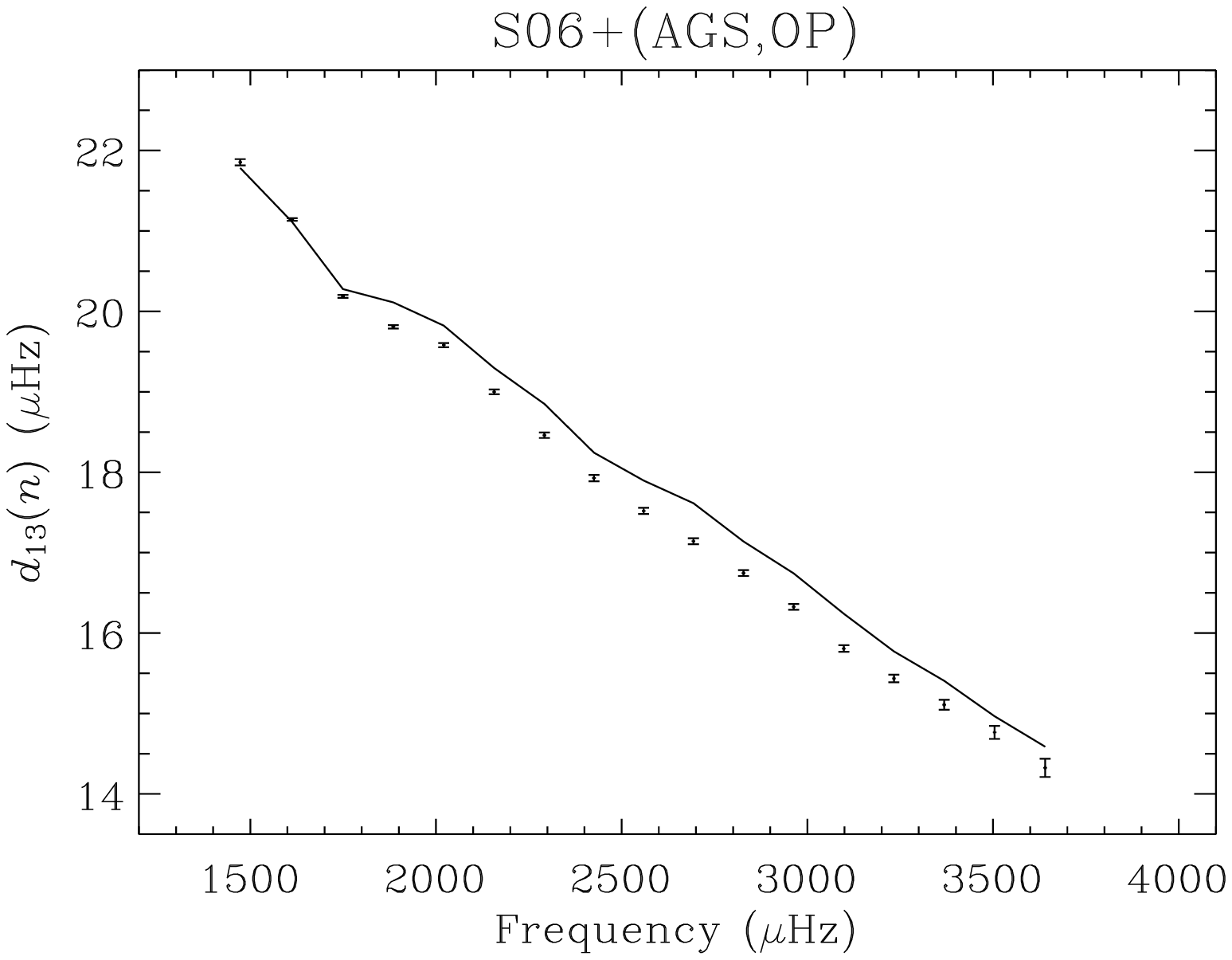}

 \caption{Solid line in each panel: fine spacings $d_{13}(n)$ of solar
 model identified in plot title. Points with error bars: $d_{13}(n)$
 from the corrected BiSON frequencies.}

 \label{fig:d1}

 \end{figure*}

\begin{figure*}

 \epsscale{1.0}
 \plottwo{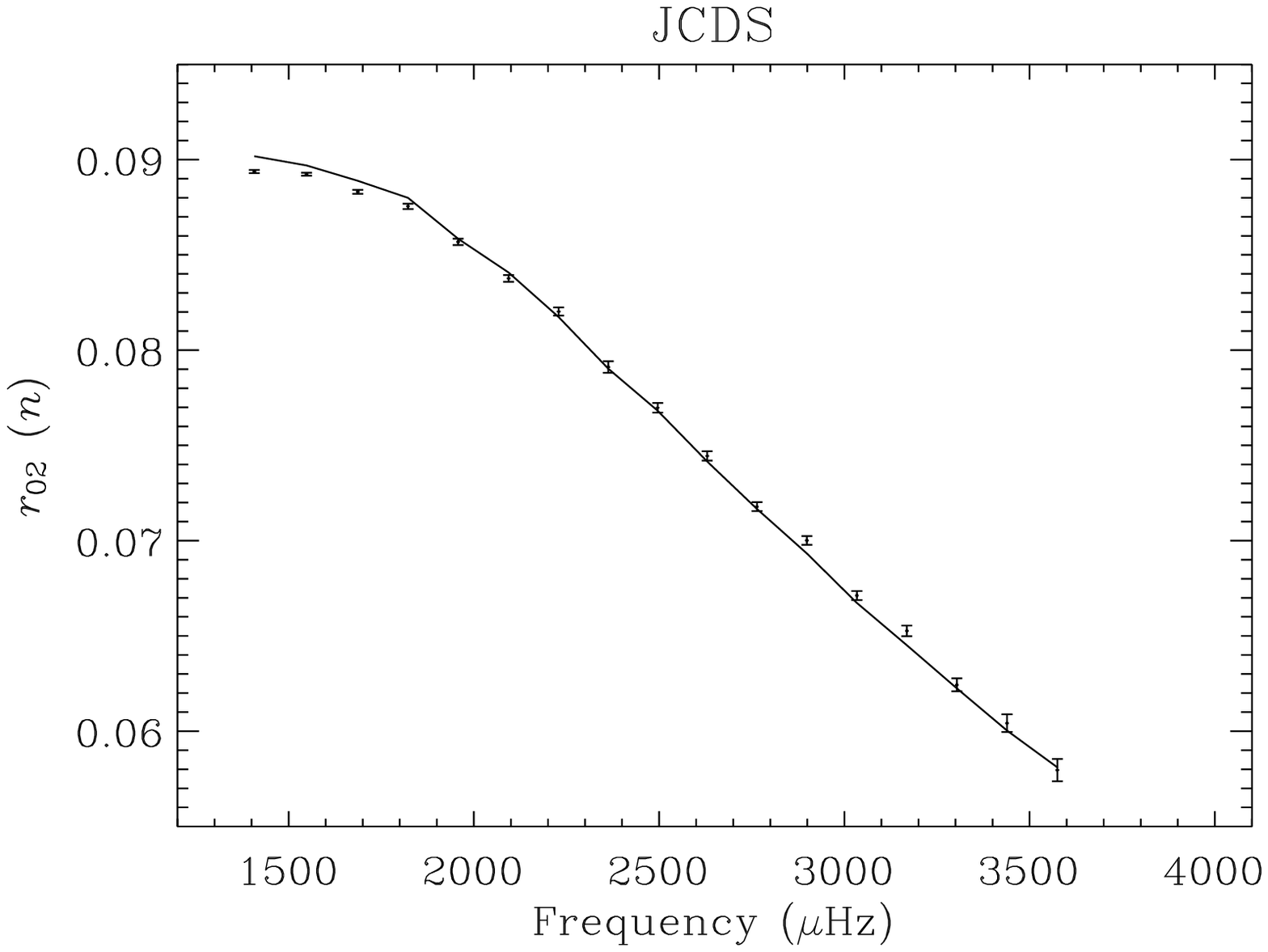}{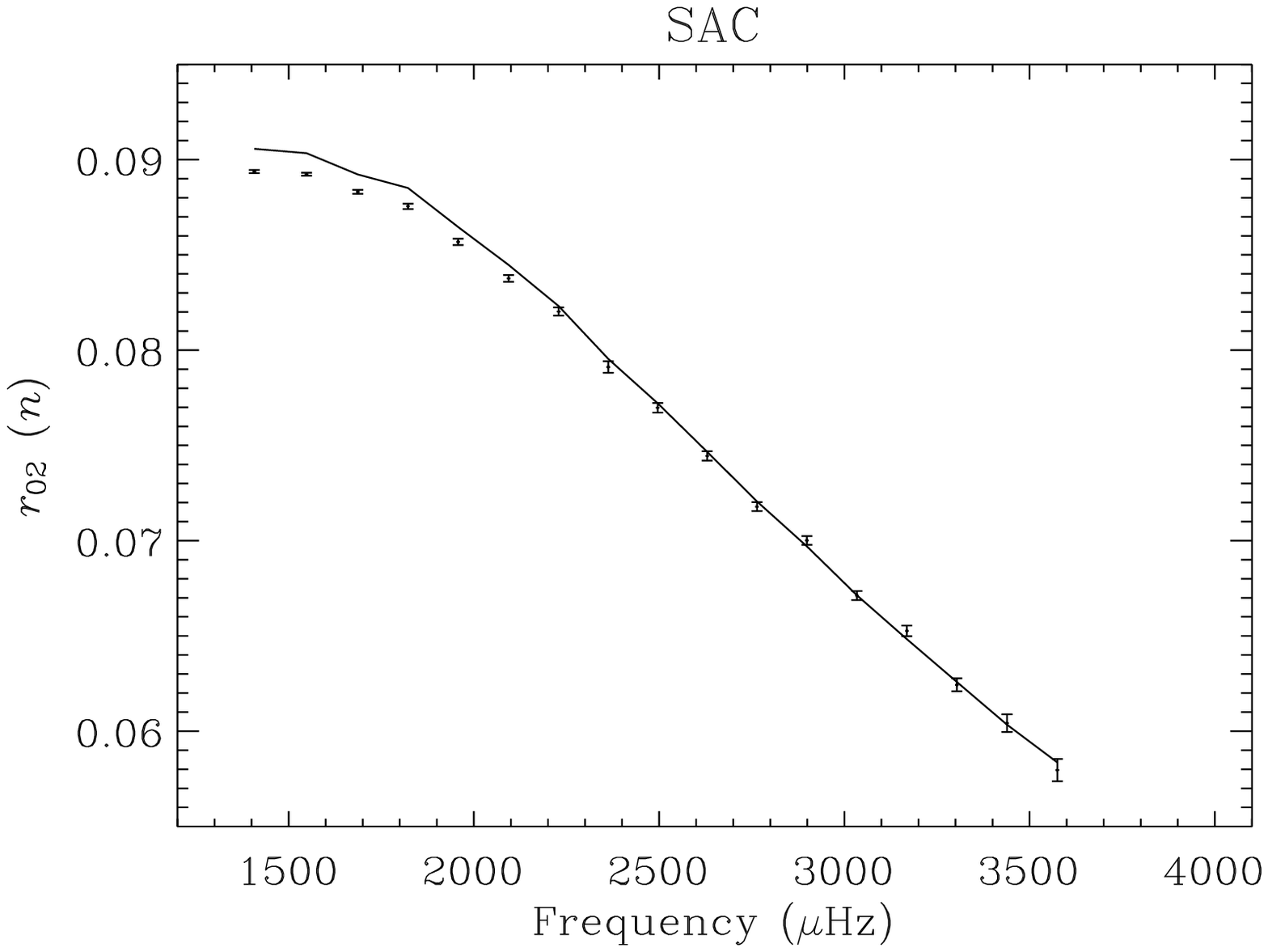}\\
 \plottwo{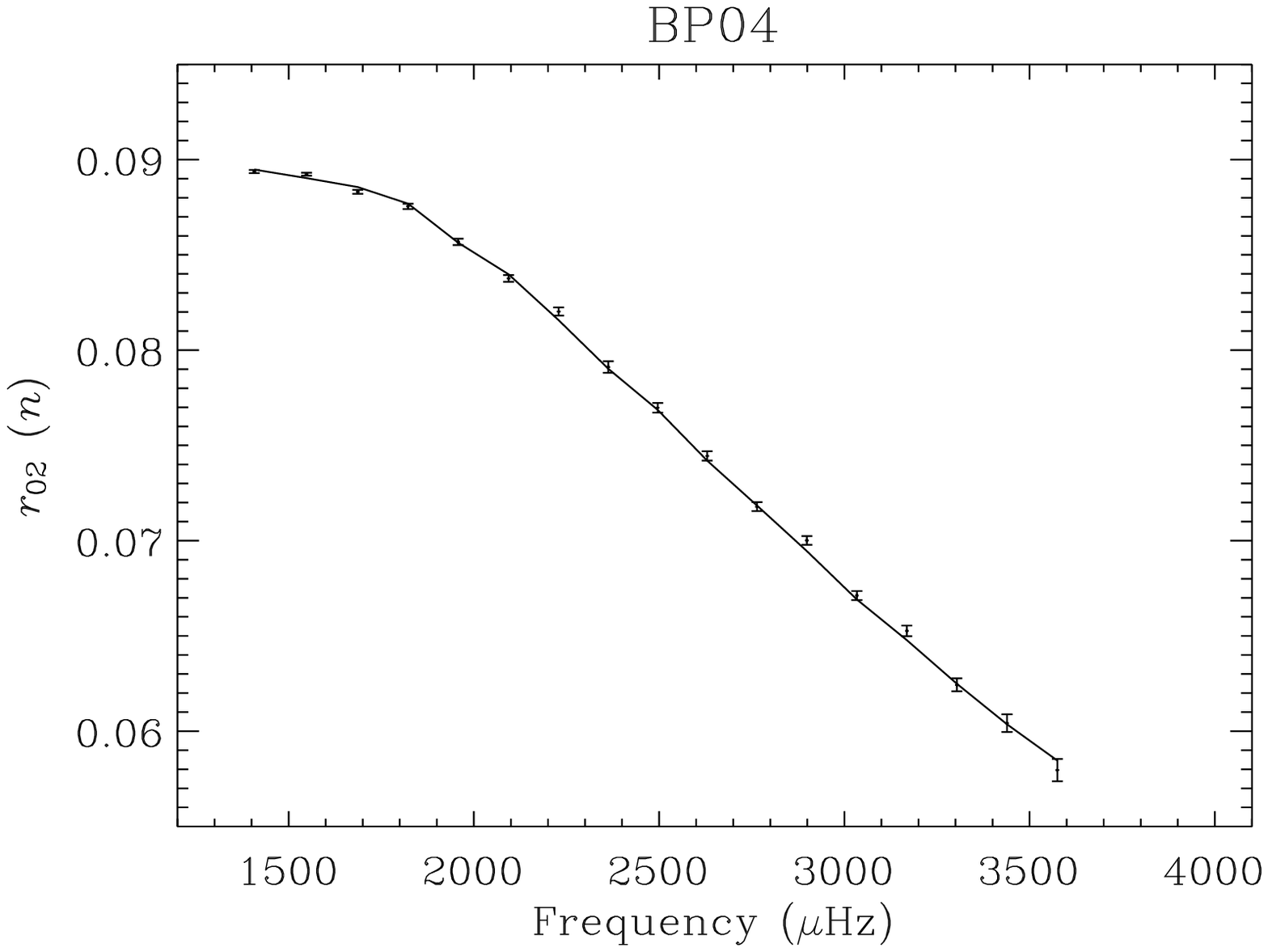}{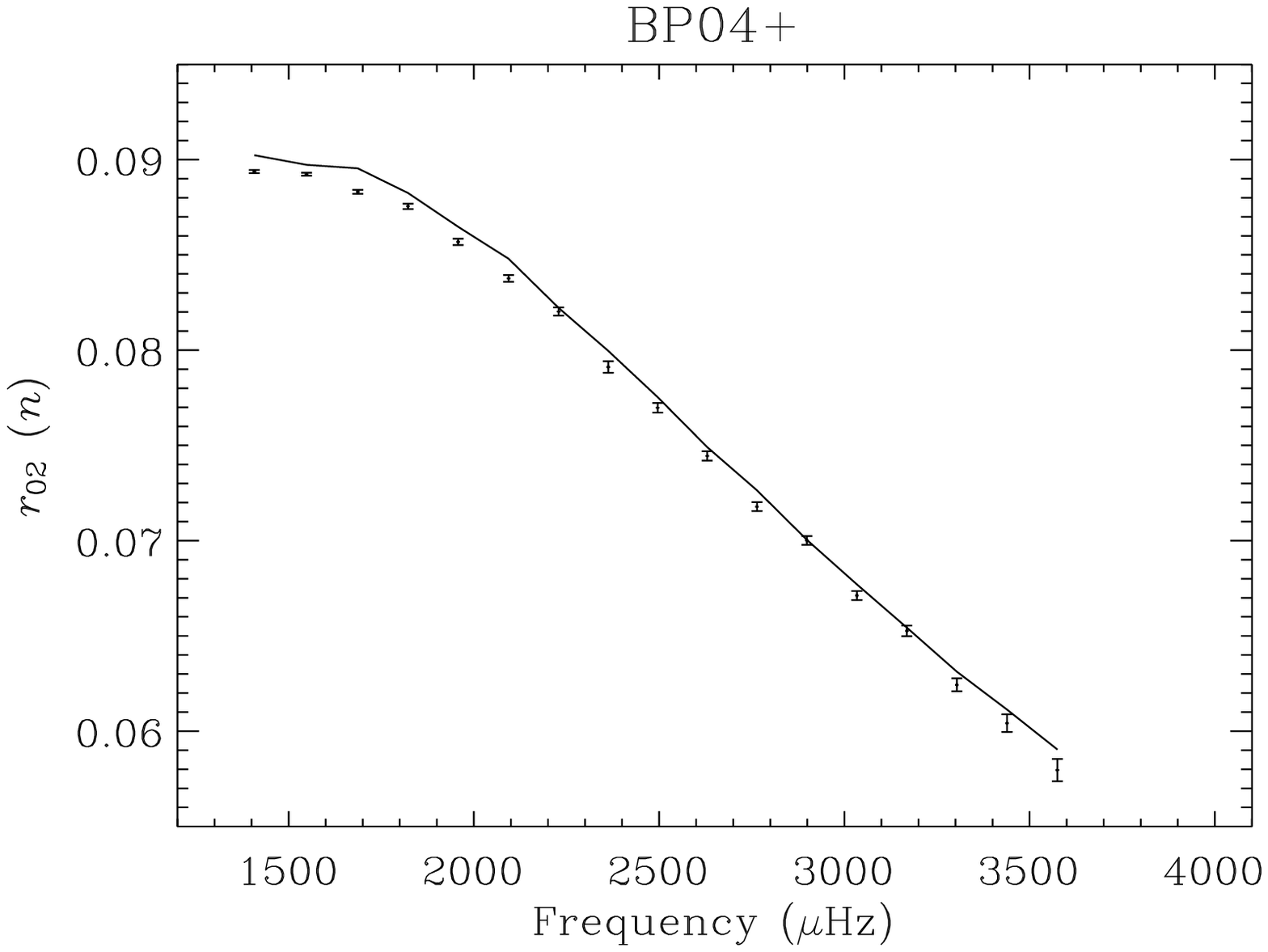}\\
 \plottwo{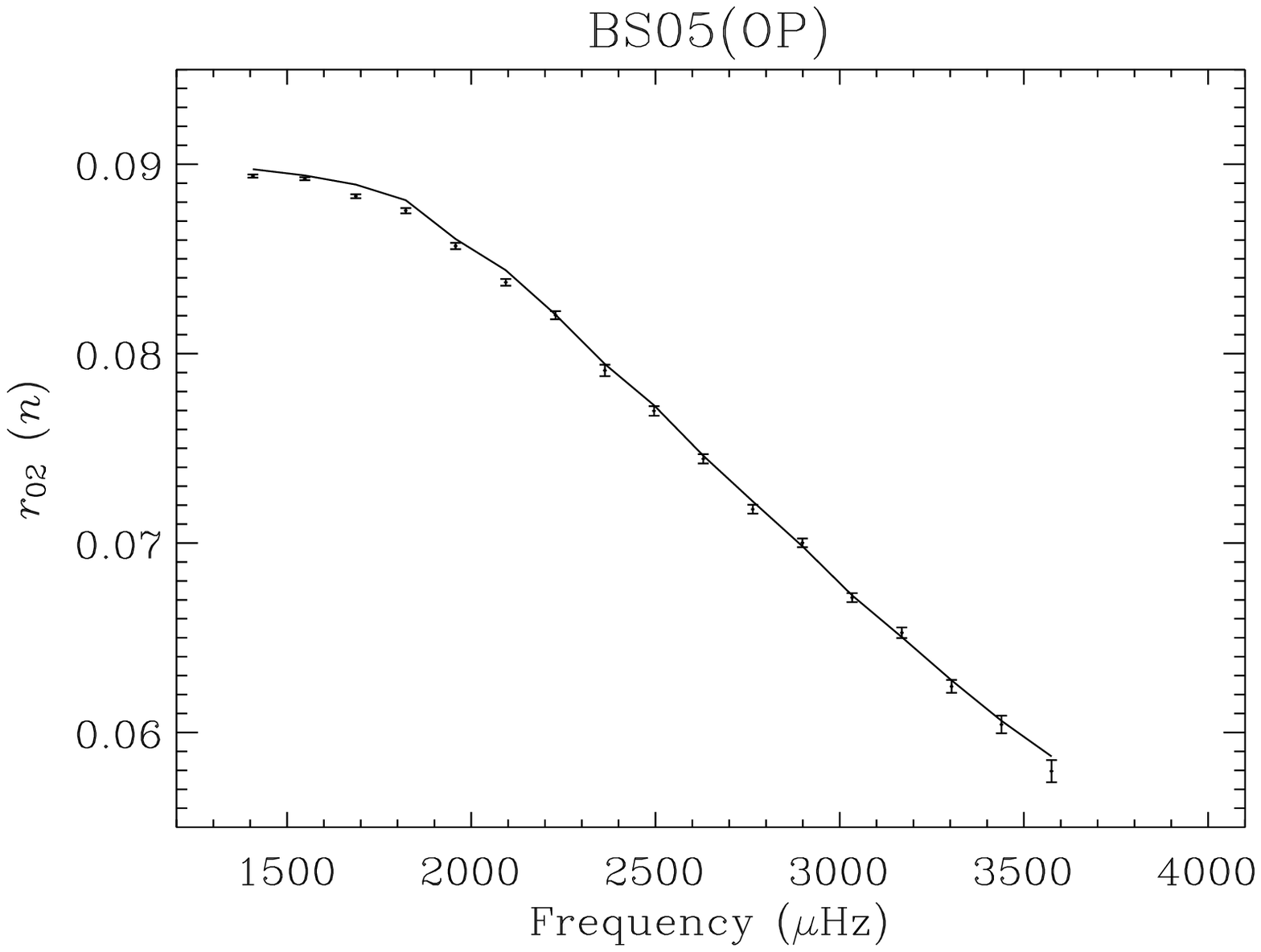}{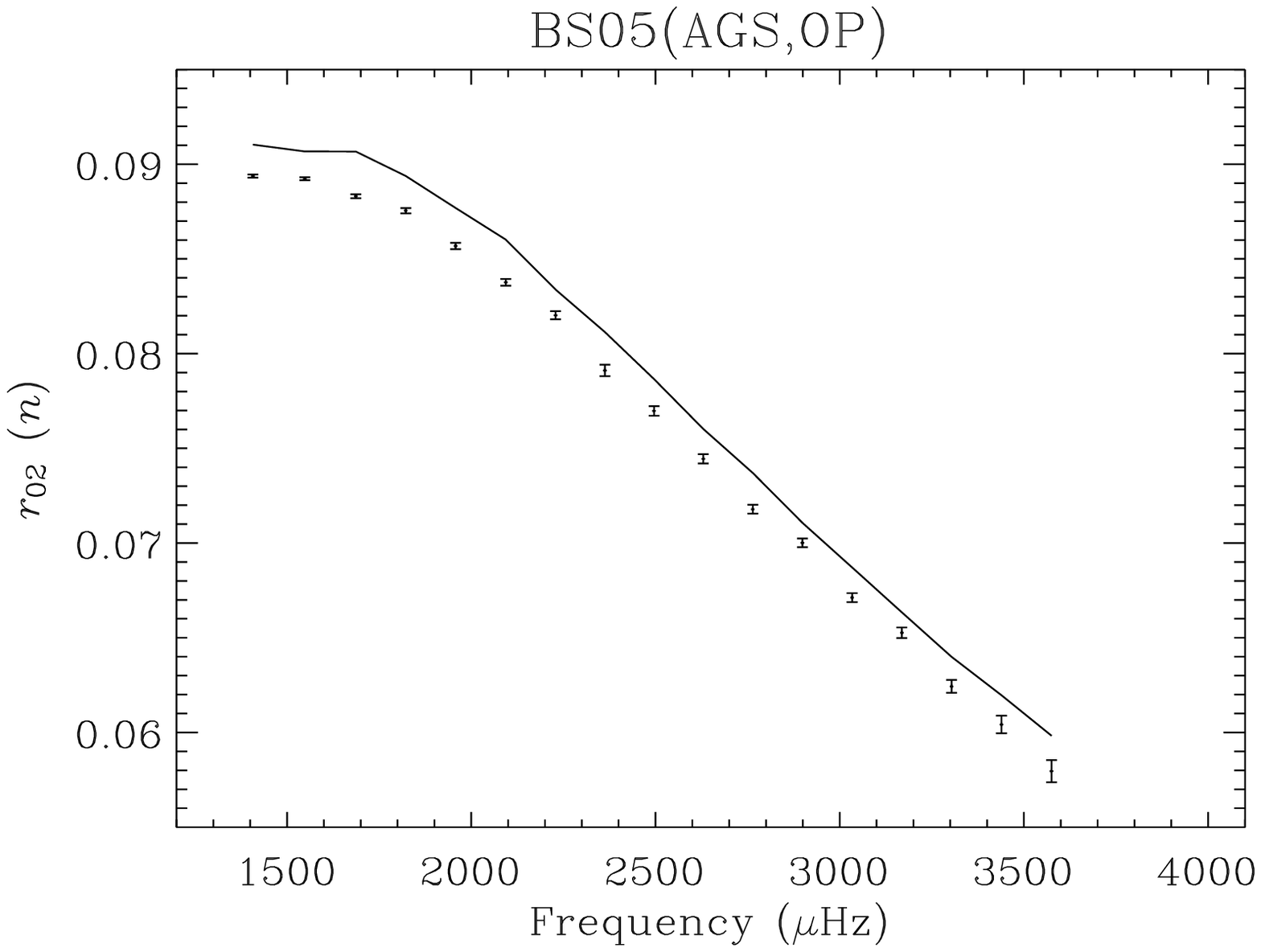}\\
 \plottwo{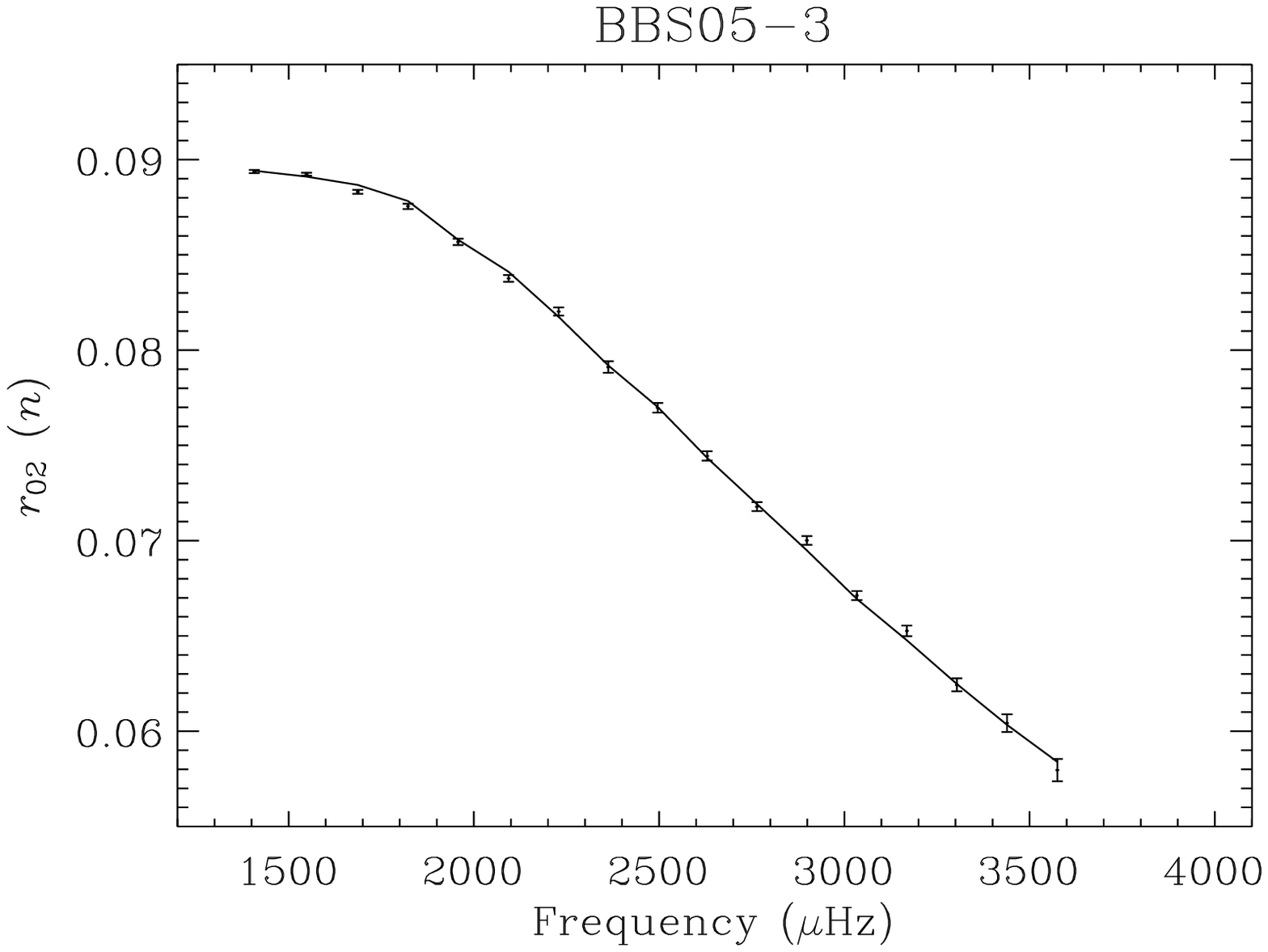}{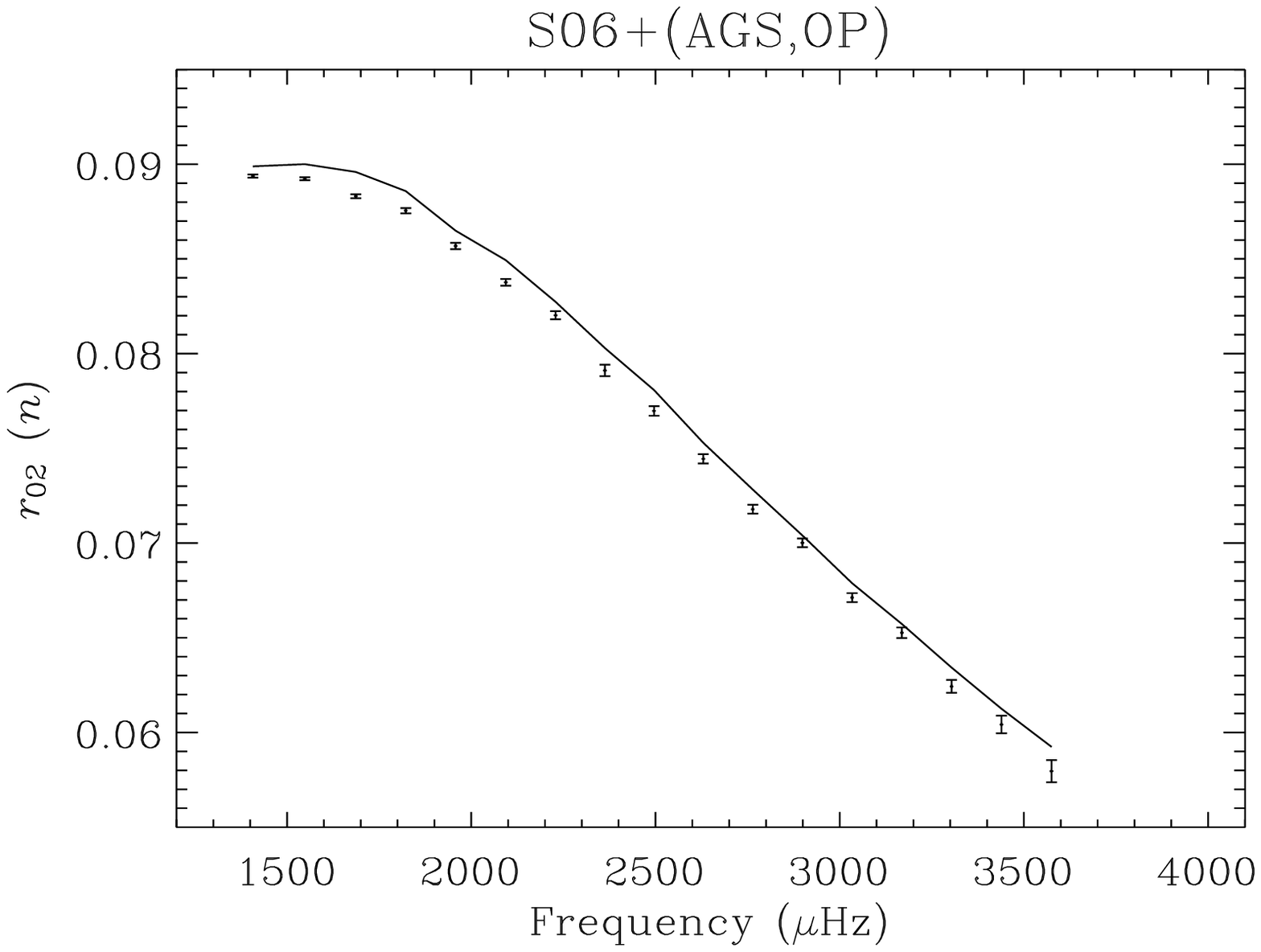}

 \caption{Solid line in each panel: separation ratios $r_{02}(n)$ of
 solar model identified in plot title. Points with error bars:
 $r_{02}(n)$ from the corrected BiSON frequencies.}

 \label{fig:sr0}

 \end{figure*}

\begin{figure*}

 \epsscale{1.0}
 \plottwo{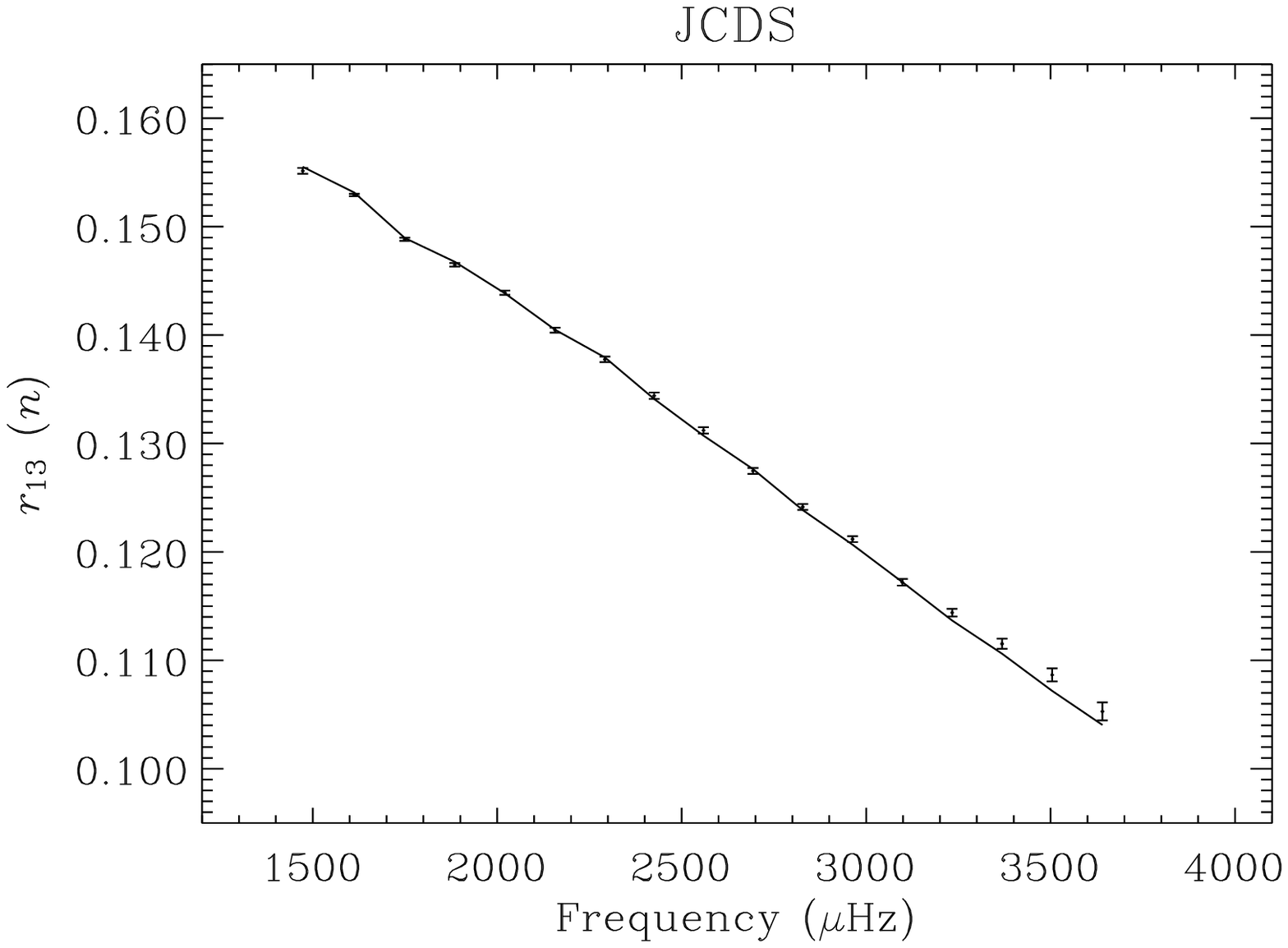}{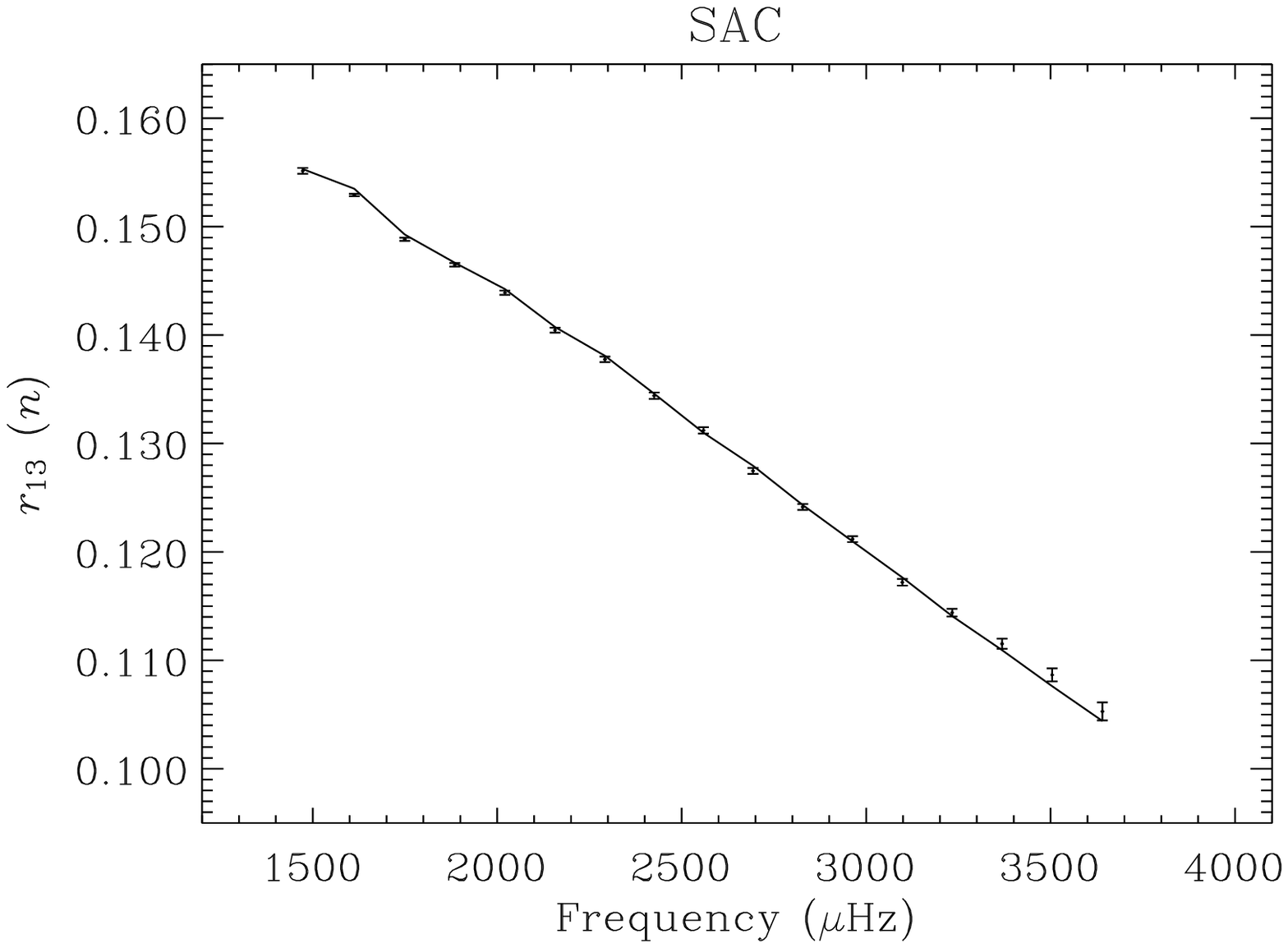}\\
 \plottwo{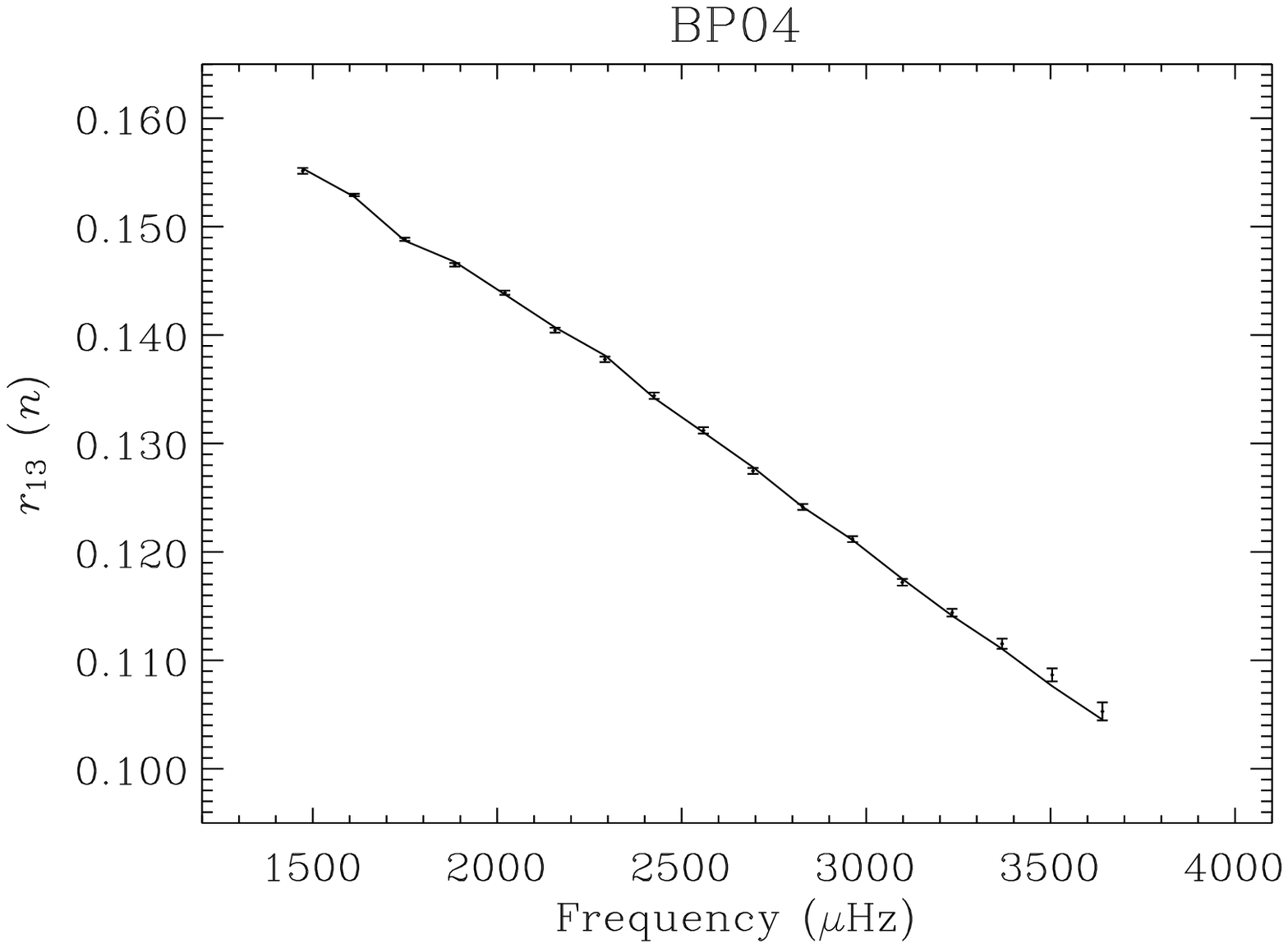}{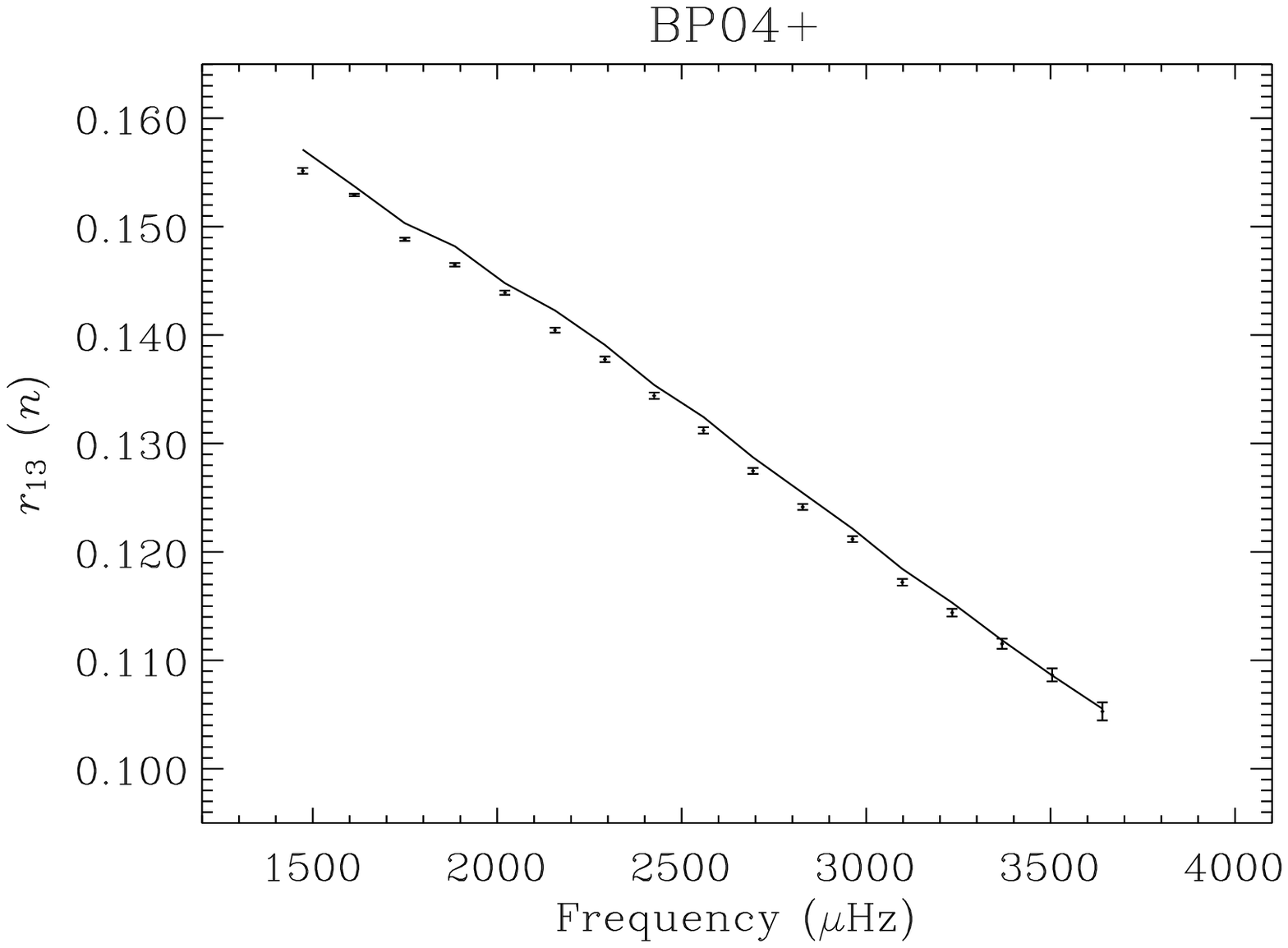}\\
 \plottwo{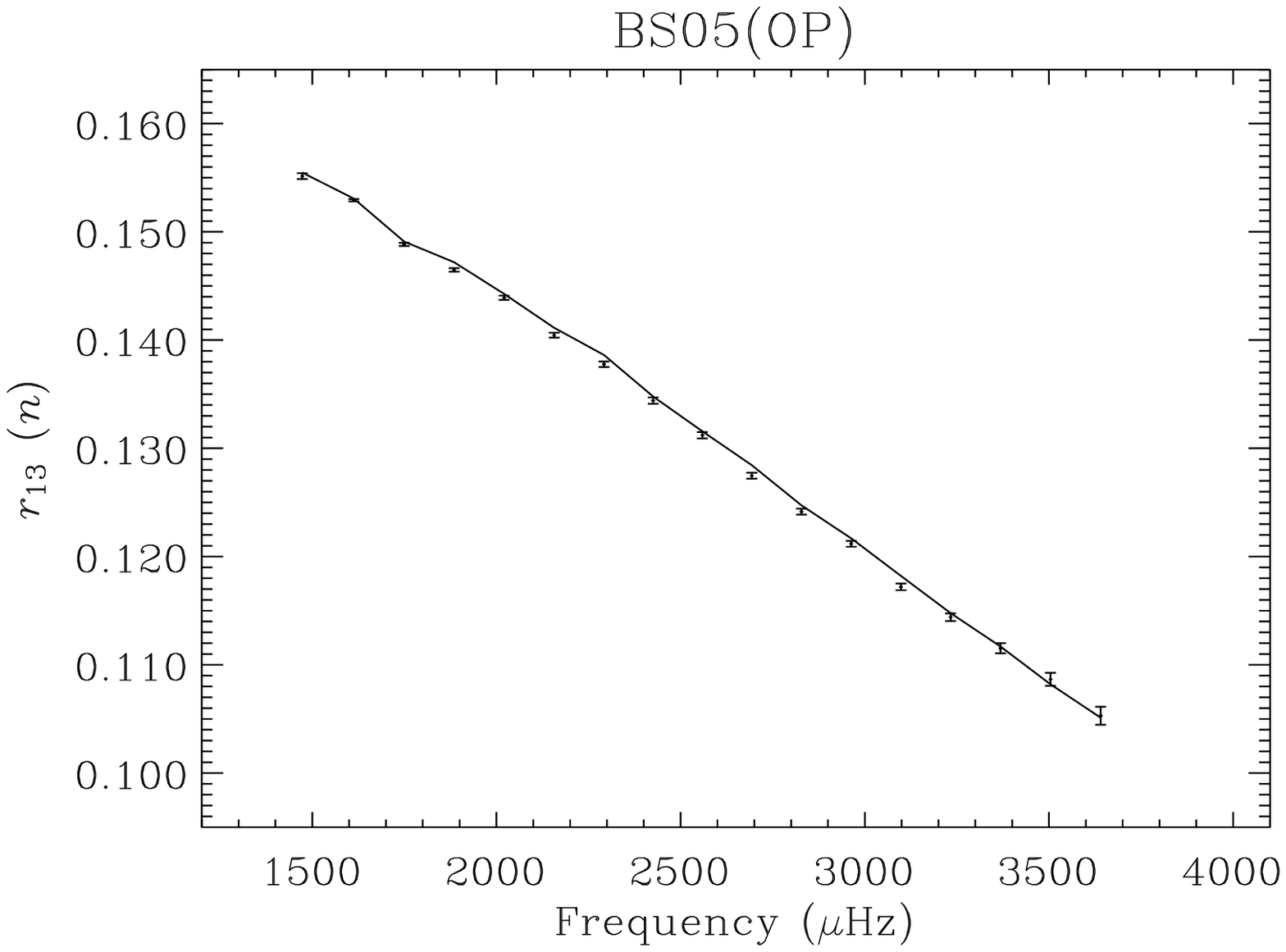}{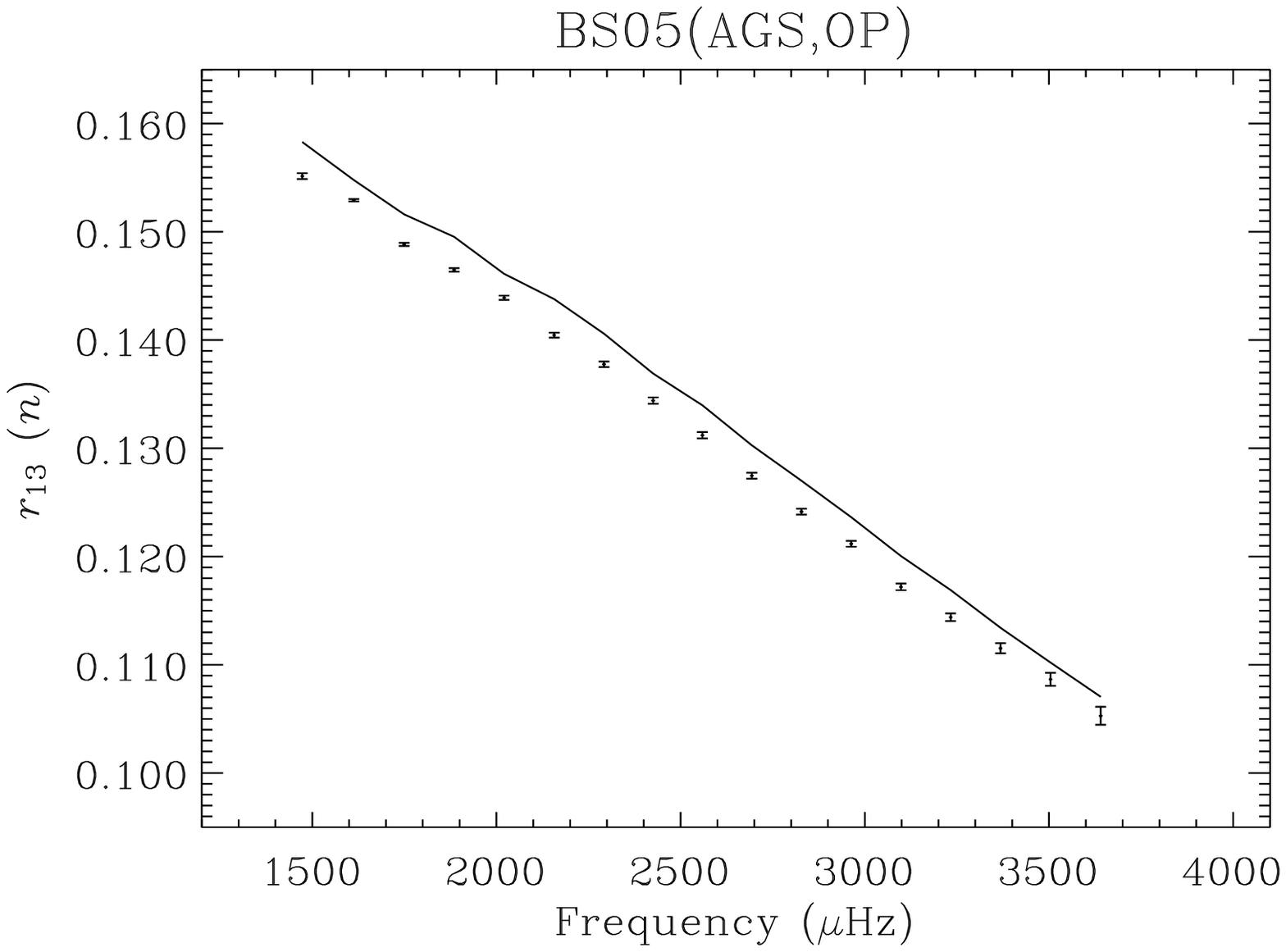}\\
 \plottwo{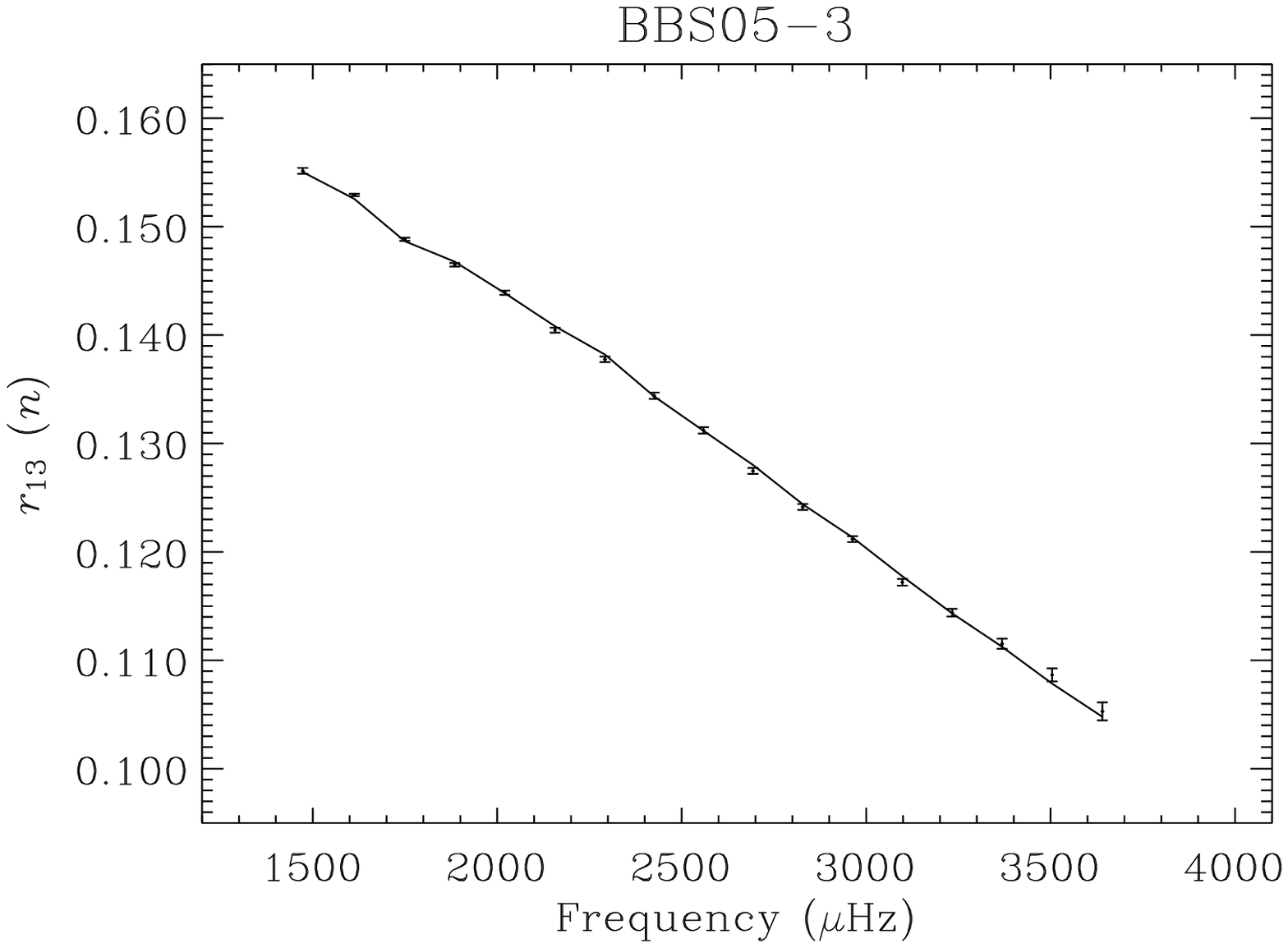}{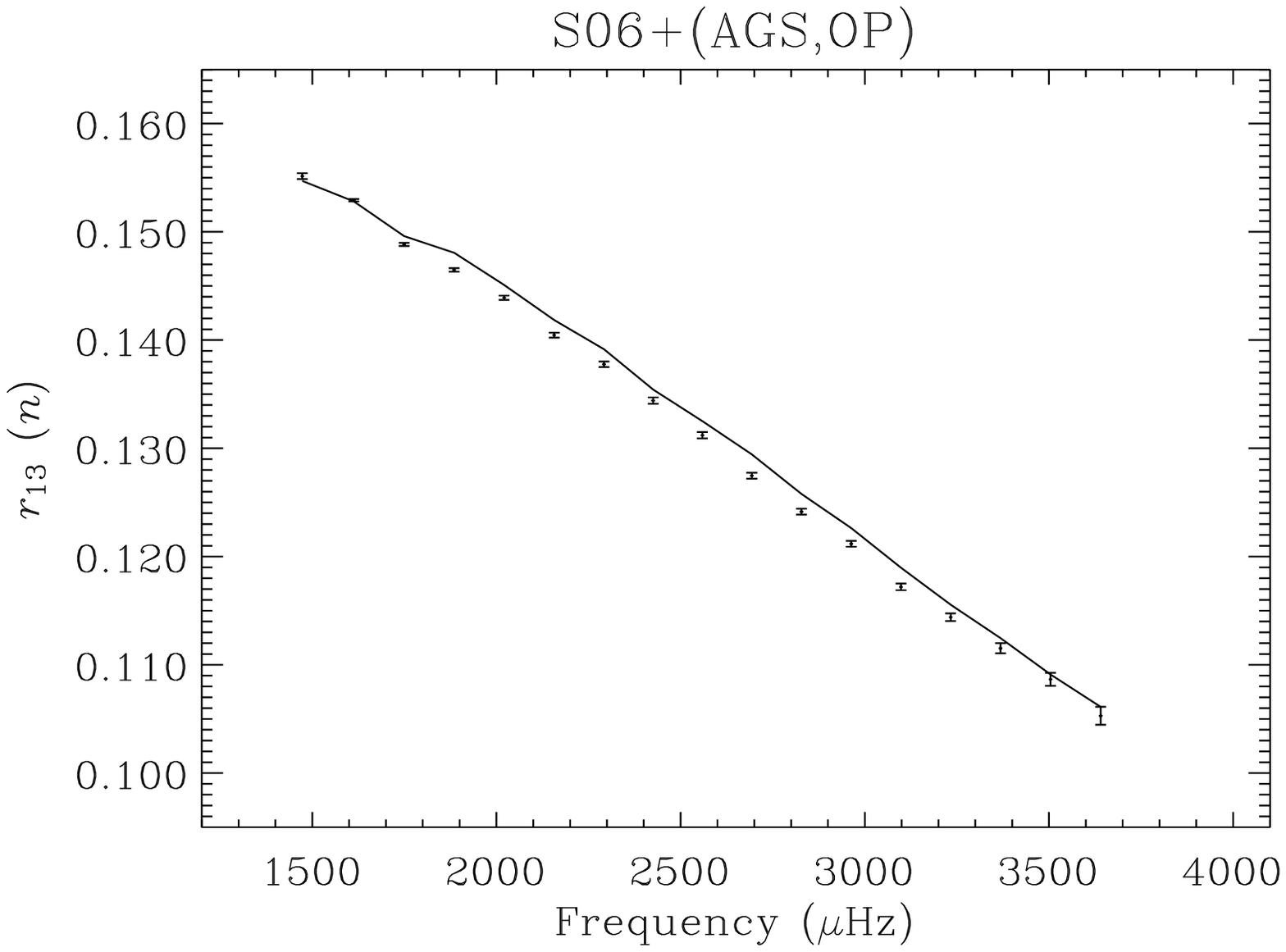}

 \caption{Solid line in each panel: separation ratios $r_{13}(n)$ of
 solar model identified in plot title. Points with error bars:
 $r_{13}(n)$ from the corrected BiSON frequencies.}

 \label{fig:sr1}

 \end{figure*}

Each panel of Fig.~\ref{fig:d0} shows the fine spacings, $d_{02}(n)$,
from one of the solar models. Model data are rendered as a solid line,
and the models are identified in the plot titles. The panels also show
for direct comparison the $d_{02}(n)$ calculated from the corrected
BiSON frequencies (points with error
bars). Figs.~\ref{fig:d1},~\ref{fig:sr0} and~\ref{fig:sr1} show
similar plots for the $d_{13}(n)$, $r_{02}(n)$ and $r_{13}(n)$ data
respectively.

\begin{figure*}

 \epsscale{1.0}
 \plotone{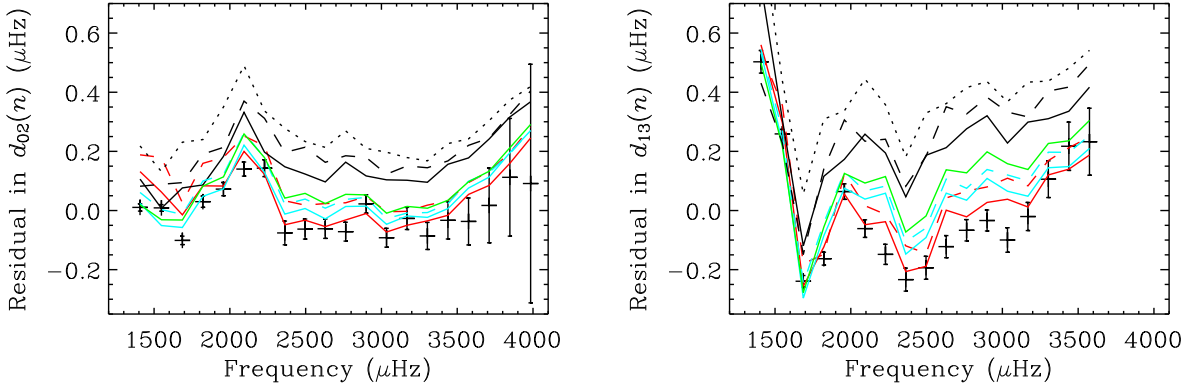}
 
 \caption{Residuals given by subtraction of best-fit straight line (to
 fine spacings versus frequency) from BiSON fine spacings and each set
 of model fine spacings. Left-hand panel: residuals in
 $d_{02}(n)$. Right-hand panel: residuals in $d_{13}(n)$. BiSON data
 are plotted with their associated error bars. Various models rendered
 as follows: JCDS (red, solid); SAC (red, dashed); BP04 (cyan, solid);
 BP04+ (black, solid); BS05(OP) (cyan, dashed); BS05(AGS,OP) (black,
 dotted); BBS05-3 (green); S06+(AGS,OP) (black, dashed).}

 \label{fig:dres}

 \end{figure*}

The fine spacings and separation ratios show a marked dependence on
mode frequency. Removal of these trends allows for a more detailed
visual comparison of differences between the BiSON and model data. In
the case of the fine spacings, we have followed an approach similar to
that given in Chaplin et al. (1997). Here, a simple linear model of
the form
 \begin{equation}
 d_{l\,l+2}(n) = c_0 + c_1 \nu
 \label{eq:makeres}
 \end{equation}
was fitted to each set of BiSON fine spacings. The Fit to the
$d_{02}(n)$ data, made over the range $\sim 1408$ to $\sim 3985\,\rm
\mu Hz$, gave best-fitting coefficients of $c_0 = 15.95 \pm 0.06\,\rm
\mu Hz$ and $c_1 = (-2.24 \pm 0.02) \times 10^{-3}$; while the fit to
the $d_{13}(n)$ data, made over the range $\sim 1473$ to $\sim
3640\,\rm \mu Hz$, gave coefficients $c_0 = 26.29 \pm 0.20\,\rm \mu
Hz$ and $c_1 = (-3.35 \pm 0.08) \times 10^{-3}$. The best-fitting
models were then subtracted from the BiSON spacings, and each set of
model spacings, to yield the residuals that are plotted in the two
panels of Figs.~\ref{fig:dres} (see caption for details). By adopting
this approach of characterizing the spacings by a simple linear model
we preserve in the residuals various features in frequency that are
common to both the observations and models.  The feature around $\sim
2000\,\rm \mu Hz$ which is present in the various curves in
Fig.~\ref{fig:dres} is the signature of the HeII ionization zone. The
presence of this feature in the data shows that the fine-spacings are
not immune to the near-surface structure.

\begin{figure*}

 \epsscale{1.0}
 \plotone{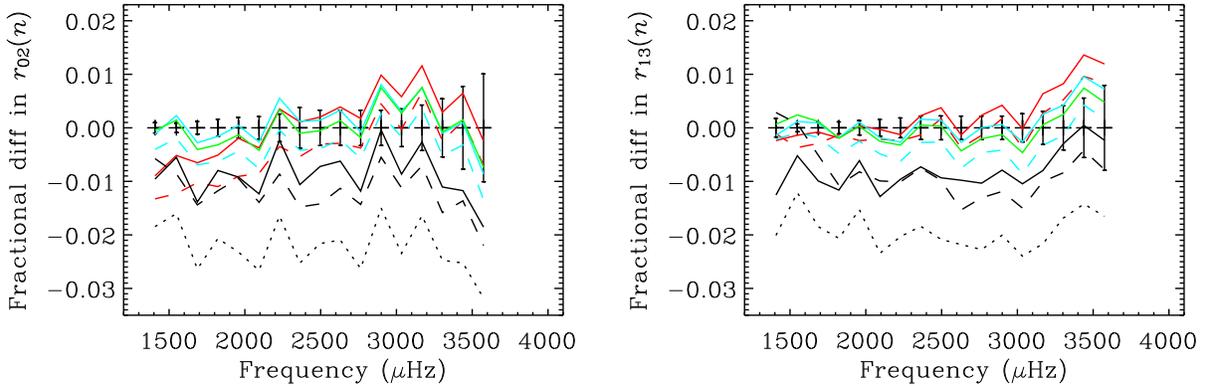}
 
 \caption{Differences between observed BiSON and model separation
 ratios (in sense BiSON minus model). Left-hand panel: differences in
 $r_{02}(n)$. Right-hand panel: differences in $r_{13}(n)$. Various
 models rendered as follows: JCDS (red, solid); SAC (red, dashed);
 BP04 (cyan, solid); BP04+ (black, solid); BS05(OP) (cyan, dashed);
 BS05(AGS,OP) (black, dotted); BBS05-3 (green); S06+(AGS,OP) (black,
 dashed).  }

 \label{fig:srres}

 \end{figure*}

The separation ratios $r_{02}(n)$ in contrast show marked departures
from linear behavior with frequency at the low-overtone end of the
plotted data. Furthermore, these data, and the $r_{13}(n)$, tend to
vary more smoothly in frequency than do their fine-spacing
counterparts. In Fig.~\ref{fig:srres} we therefore plot just the
differences between the observed and model ratios (see caption for
details).

Tables~\ref{tab:d} and~\ref{tab:sr} give quantitative measures of the
differences between the observed and model datasets. Table~\ref{tab:d}
shows weighted mean differences (in the sense BiSON minus model), and
weighted \textsc{rms} differences, for the fine spacing data (all in
$\rm \mu Hz$). The formal uncertainties on the BiSON spacings were
used to fix the weights (with the usual uncertainty-squared Gaussian
weighting applied), and to calculate an internal error on each mean or
\textsc{rms} difference. Significance levels for the differences (in
units of sigma) appear in the table in brackets, and were computed by
dividing each mean or \textsc{rms} difference by its associated
internal error.

Table~\ref{tab:sr} shows similar data for the separation ratios. Here,
the mean differences are weighted \emph{fractional} differences
between the BiSON and model values. Because each ratio is formed from
the quotient of two separations in frequency, with the uncertainties
on these separations assumed to follow Gaussian distributions,
determination of the fractional differences allows errors to be given
that are also Gaussian (and not asymmetric) in the final measure.

From inspection of the various figures, particularly
Figs.~\ref{fig:dres} and \ref{fig:srres}, and the data in the two
tables, it is apparent immediately that the models that show the
poorest correspondence with the observations are those that have
lowered abundances, i.e., models BP04+, BP05(AGS,OP) and S06+(AGS,OP).
Model BP04+ has slightly higher metallicity than models BP05(AGS,OP)
and S06+(AGS,OP), and so fares slightly better in the comparisons.
The models with higher abundances do much better: model JCDS, although
constructed with outdated inputs, agrees quite well with the
observations, as do models BP04 and BS05(OP). The non-standard model,
BBS05-3, does well too. Model SAC does not fare as well as models JCDS
or BP04, however, it still agrees better than do the models with
lower abundances. Thus we can say the models with lower than GS
abundances have core structures that do not match that of the Sun.

We shall not discuss models JCDS and SAC further, except to note again
that they fare much better than the low-$Z/X$ models. Models JCDS and
SAC were constructed using the older OPAL equation of state, which did
not treat relativistic effects at temperatures and densities relevant
to the solar core. This resulted in a somewhat deficient core
structure (see Elliot \& Kosovichev, 1998). We shall discuss in more
detail below results obtained from models that were instead
constructed with the \emph{corrected} OPAL equation of state.

Model S06+(AGS, OP) was made because a localized change in opacities
can resolve the problem of the convection zone depth in models with
AGS abundances (see, e.g. Basu \& Antia 2004; Bahcall et
al. 2005a). However, it is very clear from the fine spacings, and
separation ratios that this model fails in the core.

We find that the model BBS05-3 does reasonably.  This model was
constructed by increasing the amount of neon compared to the AGS
abundances, in addition to increasing the quantities of some of the
other elements. A strategy of increasing the amounts of neon and other
elements (within their error bars) was suggested by Antia \& Basu
(2005) and Bahcall et al. (2005) as a potential way to solve the
disagreement between the helioseismic observations and the models with
AGS abundances. Neon contributes to the opacity near the
convection-zone base. An increase in the neon abundance therefore
deepens the convection zone, bringing the models into agreement with
the helioseismically determined convection-zone depth. It is worth
adding that because of the abundance increases in some of the elements
this model has a relatively high $Z/X$ (0.02069).

Differences between the results of the various models ultimately have
to be understood in terms of the sound-speed gradient in the core
which, \emph{as per} Eq.~\ref{eq:smallsep}, determines the fine
spacings and therefore also the separation ratios. Sound speed is
determined both by temperature $T$ and mean molecular weight $\mu$,
with $c^2\propto T/\mu$.  The gradient of sound speed is therefore
determined by the gradients of temperature and $\mu$ as well.
Although the sound-speed gradient is negative in the bulk of the Sun,
it is positive at very small radii because of the different radial
dependences of $T$ and $\mu$, with the increase in $\mu$ at smaller
$r$ compensating the increasing $T$ to give a lower $c$.

\begin{figure*}

 \epsscale{1.0}
 \plottwo{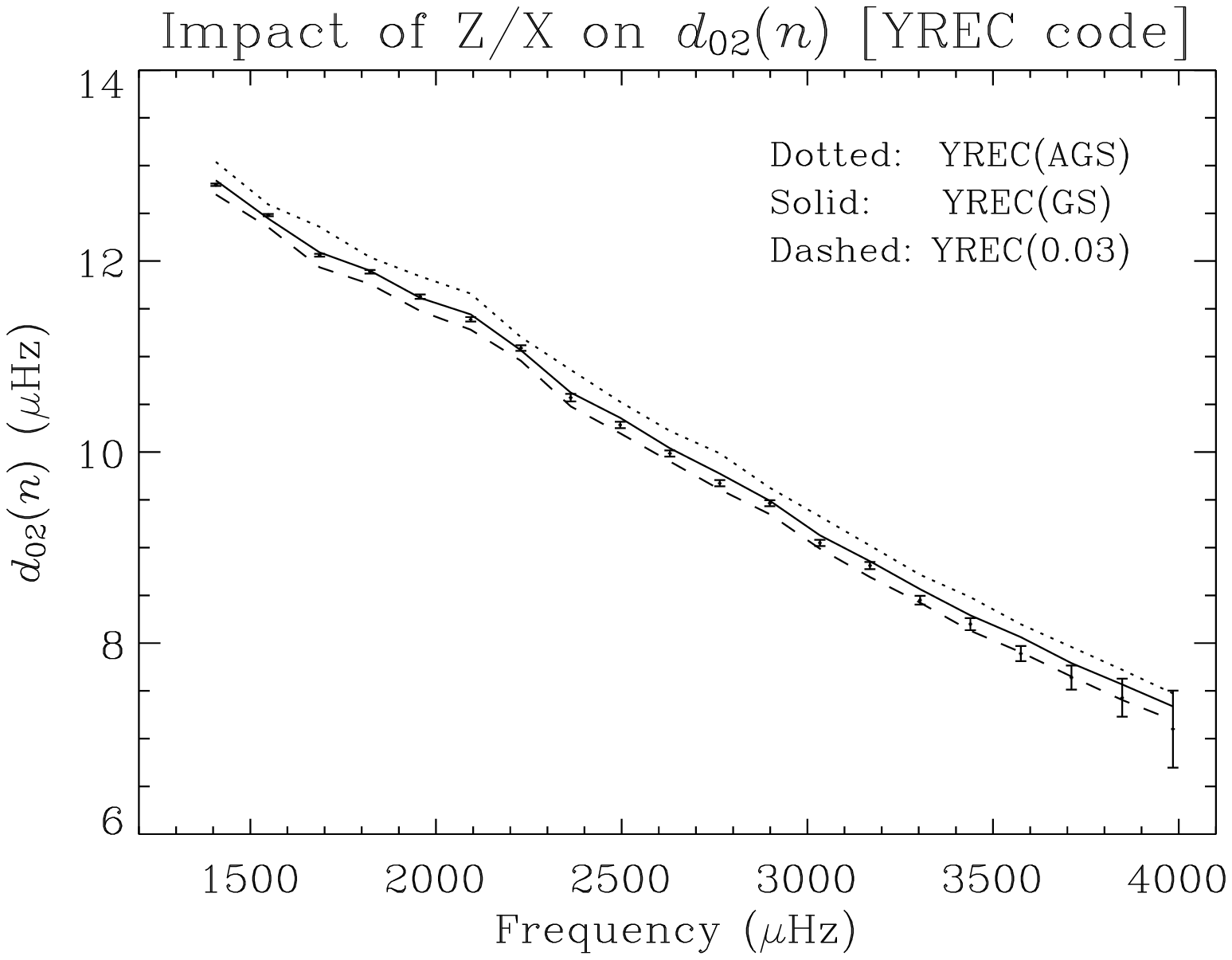}{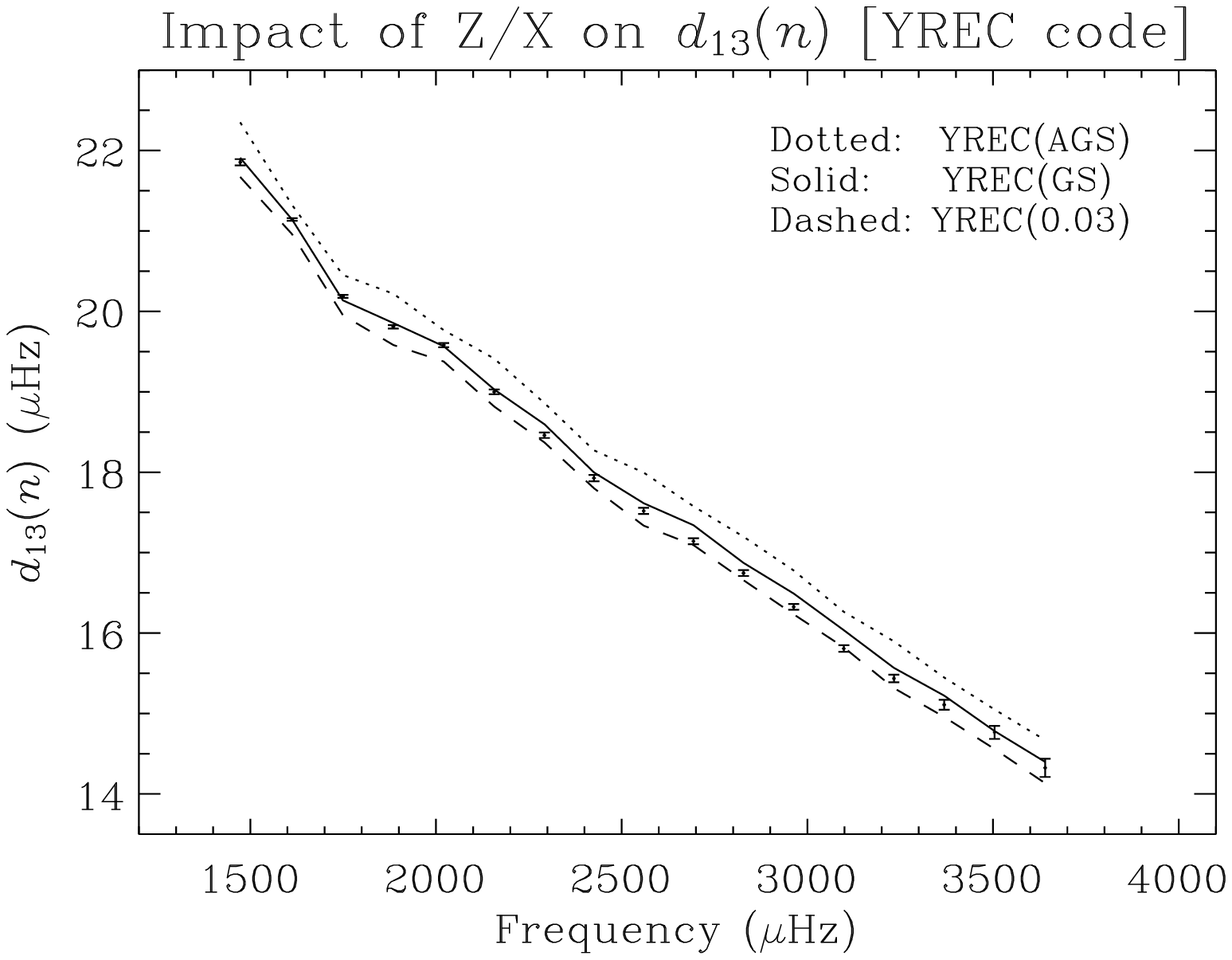}\\
 \plottwo{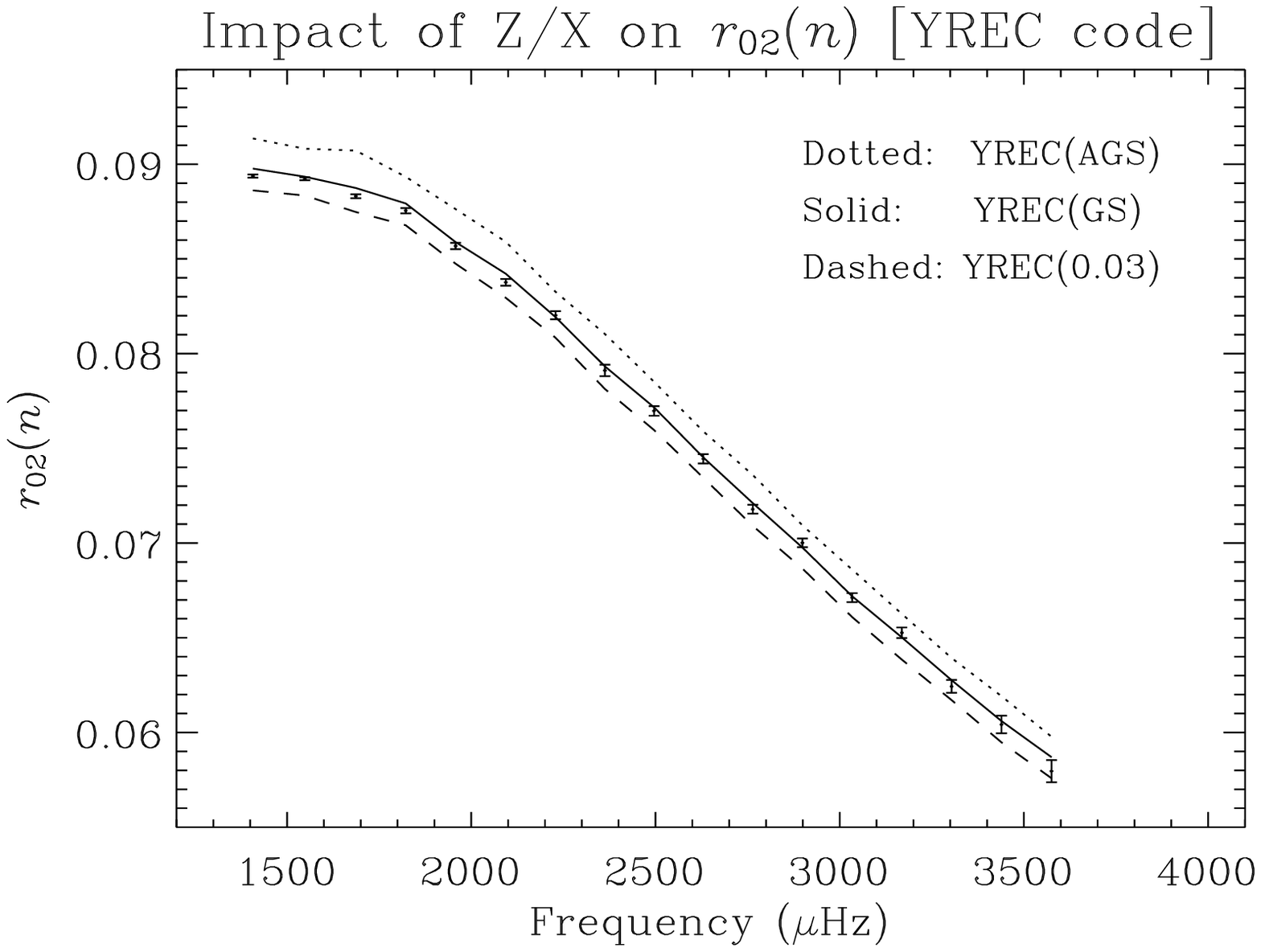}{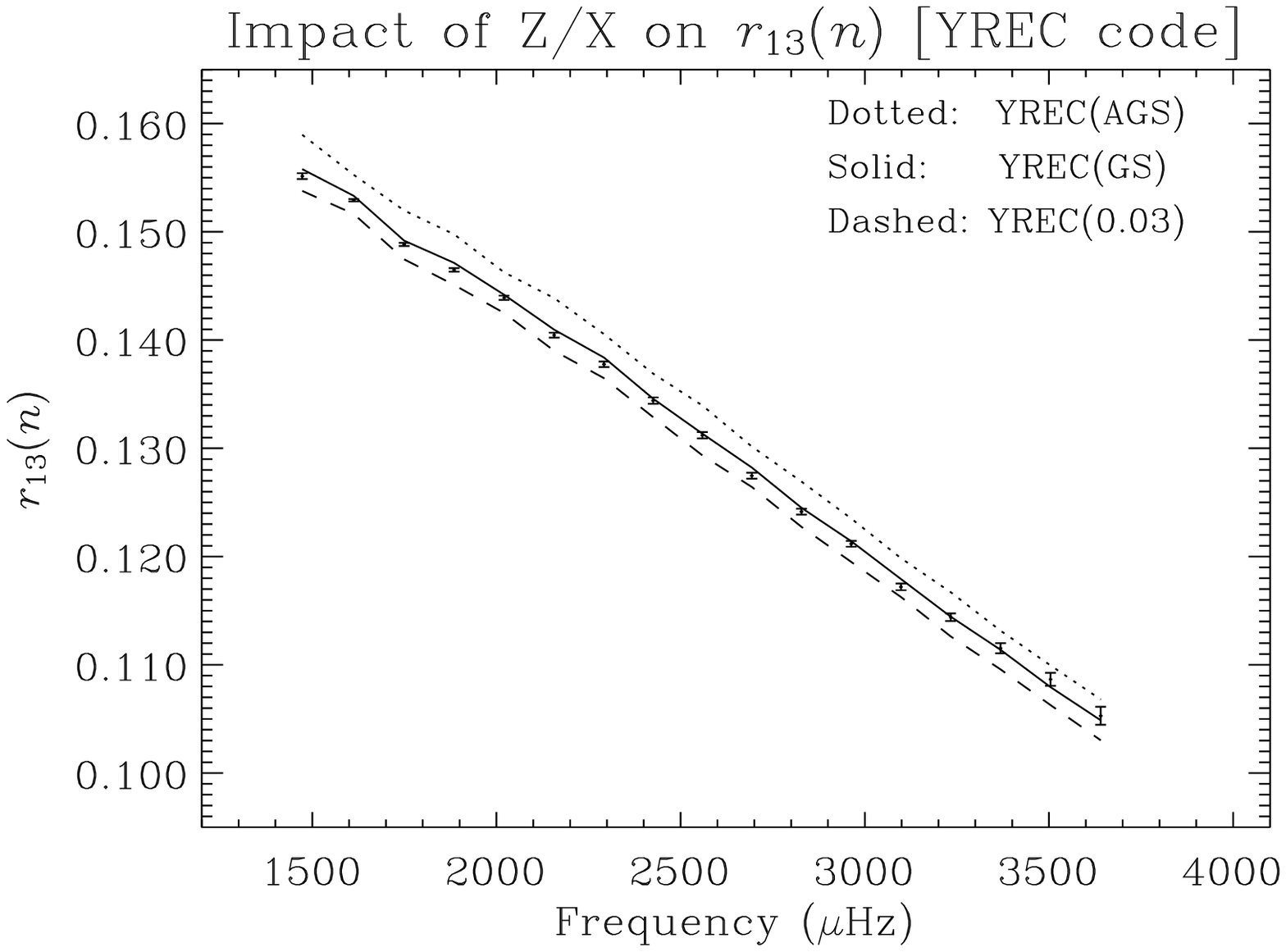}

 \caption{Impact of changes to $Z/X$ in the models. The upper panels
 show the fine spacings of three models computed with the YREC
 evolution code and the OPAL equation of state, but with different
 $Z/X$. Solid line: model YREC(GS), with $Z/X=0.0229$. Dotted line:
 model YREC(AGS), with $Z/X=0.0165$. Dashed line: model YREC(0.03),
 with $Z/X=0.03$. Lower panels: same, but for the separation ratios.}

 \label{fig:zdiff}

 \end{figure*}

\begin{figure*}

 \epsscale{1.0}
 \plottwo{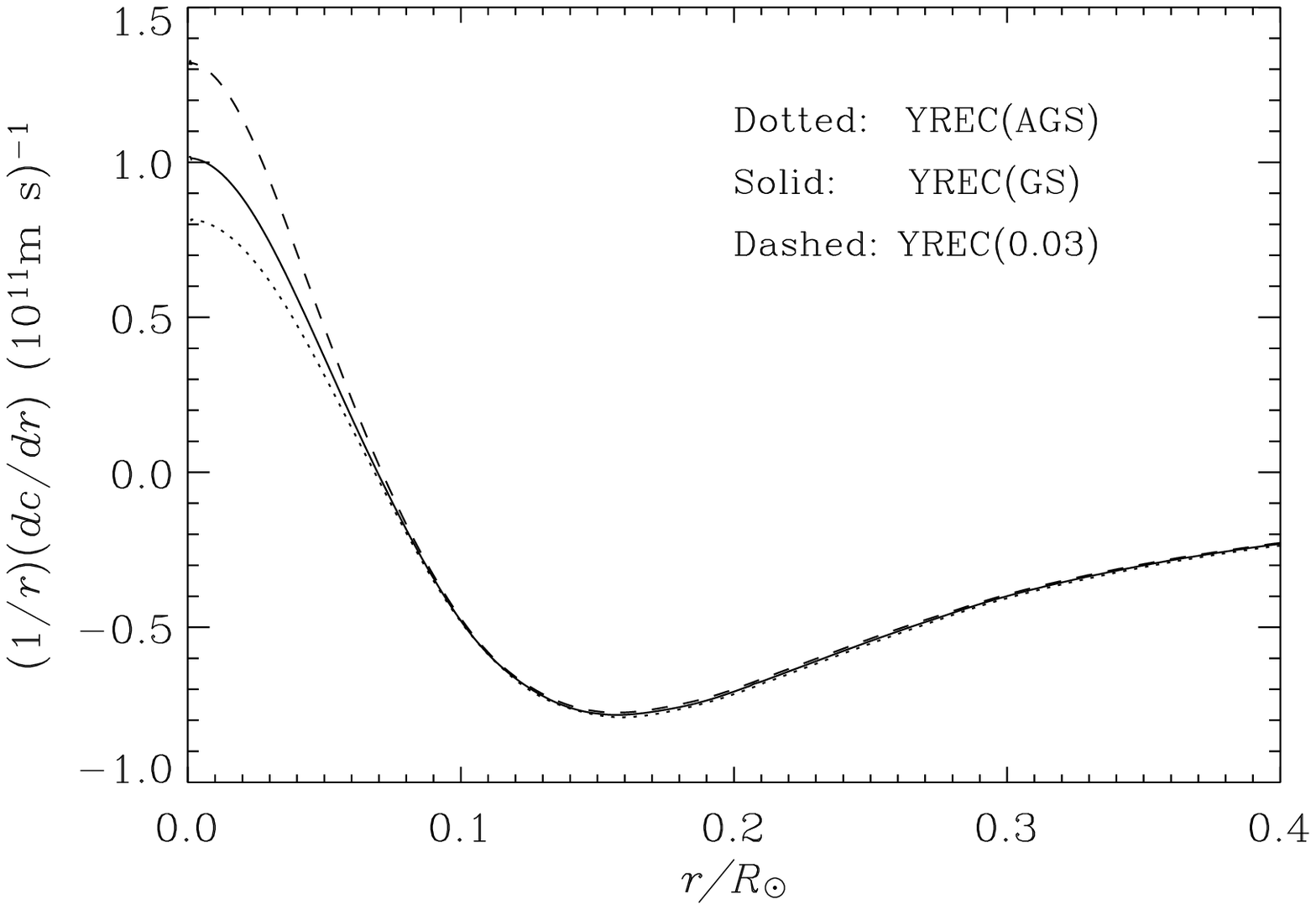}{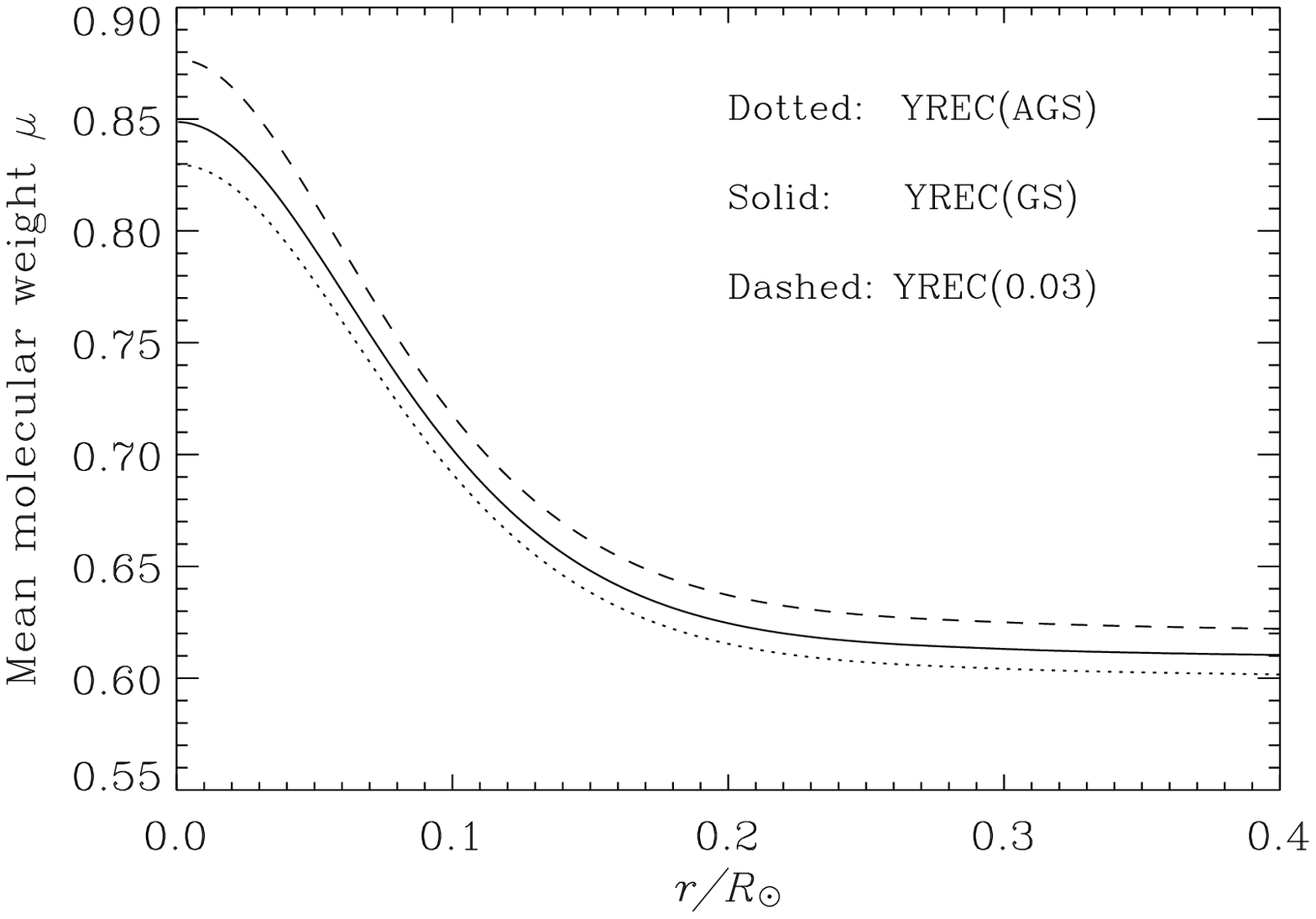}

\caption{Left-hand panel: The function $(1/r)(dc/dr)$ for the three
YREC models. This function determines to a large extent the fine
spacings, and therefore also the separation ratios. Right-hand panel:
The mean-molecular weight profile of the three YREC models.}

\label{fig:cgrad}
\end{figure*}

We can use the three YREC models to determine why the other models
with lower metallicity fare worse than models with higher metallicity
in comparison with the BiSON data. Fig.~\ref{fig:zdiff} shows the fine
spacings and separation ratios of the YREC models, together with the
BiSON data (points with error bars). We can see that increasing the
surface $Z/X$ decreases the fine spacings and separation ratios. To
explain these results we show in Fig.~\ref{fig:cgrad} the sound-speed
gradient and the mean molecular weight profiles of the YREC models.
It is clear that differences between the models lie very close to the
center, where $dc/dr$ is positive.  A more detailed analysis of the
structure of the YREC models shows that the $-{\sqrt{T/\mu^3}}\;(d\mu/dr)$ term
in the sound-speed gradient is instrumental in causing the
differences, in particular the factor $\sqrt{1/\mu^3}$. The models have very
similar temperatures, temperature gradients and $\mu$ gradients, but
the larger $\mu$ in models with higher $Z/X$ increases $dc/dr$ in the
core, thereby decreasing the value of the integral in
Eq.~\ref{eq:smallsep}.

The $\mu$ dependence of the sound-speed gradient explains why models
BP04+ and BBS05-3 do not behave like BS05(AGS, OP) and S06+(AGS,OP).
Model BP04+ does not have as low metallicities as BS05(AGS, OP) and
S06+(AGS,OP) and hence fares better than these models. That said, the
metallicity of BP04+ is low enough to give larger fine spacings and
separation ratios than are seen in the GS models.  

Model S06+(AGS, OP) does not do well in the core, despite satisfying
the helioseismic constraints at the base of the convection zone,
because $\mu$ in its core remains  similar to that in model
BS05(AGS, OP). The change in opacity in S06+(AGS, OP) does however
induce changes in the rest of the model (solar structure is, after
all, determined by a set of highly coupled equations), and this causes
enough difference in the core temperature gradient to give different
fine spacings and separation ratios compared with BS05(AGS, OP).

This leads us to the question of the differences in the spacings and
ratios of models BP04 and BS05(OP).  That there are differences is not
surprising because even though they have the same metallicity, the
models were constructed with slightly different opacities. The
temperature gradient in the radiative zone (and that includes the
solar core) is determined by the opacity, and hence different
opacities will imply a different temperature gradient.  Differences in
the treatment of diffusion and $^{17}O$ burning also have a small
effect because both change the $\mu$ gradient slightly. Nonetheless,
the overall opacity differences in the models lead to only modest
differences in the fine spacings and separation ratios.

From the above discussion it is clear that models with low
metallicities, such as those with AGS abundances, do not satisfy the
observational constraints imposed by the BiSON fine-spacing and
separation-ratio data. Though subtle changes in physics (such as in
the formulation of diffusion, etc.)  do change the core structure, the
changes are sufficiently small to allow us to distinguish between the
GS and AGS abundances. In fact, looking at Fig.~\ref{fig:zdiff}, we
might be tempted to determine the solar metallicity using the fine
spacings and separation ratios.

\section{Implications for neutrino uncertainties}
\label{sec:neut}

Uncertainties in solar neutrino predictions arise from uncertainties
in inputs to the solar models. It has, however, been long recognized
that the predominant source of uncertainty is the composition (Sears
1964, Bahcall 1966).  As a result there has been a large body of work
devoted to evaluating neutrino uncertainties, among the more recent
being Fiorentini \& Ricci (2002), Couvidat et al. (2003), Boothroyd \&
Sackmann (2003), Bahcall \& Pinsonneault (2004), Young \& Arnett
(2005), Bahcall \& Serenelli (2005), and Bahcall et al. (2006).

Bahcall \& Serenelli (2005) performed a very detailed and rigorous
investigation of how the abundances of individual elements affect
solar neutrino predictions.  Their study was motivated by the fact
that models with AGS abundances show discrepancies with respect to the
Sun just below the solar convection zone.  Bahcall \& Serenelli (2005)
found that the largest uncertainties in the $^7Be$ and $^8B$ neutrinos
are due to the uncertainty of the solar iron abundance. However,
uncertainties in oxygen, neon, silicon and sulphur also contribute
significantly to the $^7Be$ and $^8B$ flux uncertainties. 
The p-p neutrino flux is most
sensitive the changes in the iron abundance, but because carbon 
uncertainties are larger, both elements
dominate and contribute comparable amounts to the total uncertainty.
It should be noted that the $^8B$ neutrinos are the ones detected by
the water (Kamiokande \& Superkamiokande) and heavy-water detectors
(Sudbury Neutrino Observatory).  $^8B$, $^7Be$ and pep neutrinos are
detected by Chlorine detectors. In order to calculate the uncertainty,
Bahcall \& Serenelli (2005) adopted what they called a
``conservative'' uncertainty in the abundances. This uncertainty is
basically the difference between the GS and AGS abundance values.
Their so-called ``optimistic'' uncertainty was based on the error bars
on the abundance from the AGS table. The ``conservative''
uncertainties are much larger than the ``optimistic'' uncertainties.
It should be noted that quoted uncertainties in the GS and AGS tables
are fairly similar. Work done by Bahcall et al. (2006) confirmed these
results through an elaborate Monte-Carlo simulation.

Our current work has shown that models with AGS abundances fare much
worse than models with GS abundances. We also see, from
Fig.~\ref{fig:zdiff}, that the uncertainties in the abundances are
probably much less than the difference between the GS and AGS
values. We therefore conclude that the ``conservative'' uncertainties
in the neutrino fluxes are unduly pessimistic and can be lowered
substantially. Not having done the full analysis, we are unable to say
if we are justified in lowering the levels to the ``optimistic''
levels of Bahcall \& Serenelli (2005) and Bahcall et al. (2006).

Our conclusion however comes with one condition, which is related to
the model with modified neon abundance (BBS05-3) and the fact it did
well in comparisons with the BiSON data. The solar neon abundance is
very uncertain since the abundance cannot be measured at the
photosphere.  This had led Antia \& Basu (2005) and Bahcall et
al. (2005b) to propose increasing the neon abundance as a solution to
the helioseismic problems posed by the AGS models.  Drake \& Testa
(2005) found that most neighboring stars seem to have a much higher
Ne/O ratio compared to the Sun, which supported increasing the neon
abundance. However, Schmelz et al.~(2005) and Young (2005) each
reanalyzed solar X-ray and UV data and found that the Ne/O ratio of
the Sun is indeed low.  Bochsler et al. (2006) also suggest a low
solar Ne/O abundance.  A more detailed analysis of the convection zone
also points to the fact that an enhanced neon abundance will not
reconcile models with the helioseismic data (Basu \& Antia 2006). Thus
although the neon issue has not been resolved completely, it looks
increasingly unlikely that the solar neon to oxygen ratio is high. We are
therefore reasonably confident about our claim that uncertainties in
the predicted solar neutrino fluxes can be lowered.

\section{Conclusions}
\label{sec:conclu}

We have used BiSON data collected over 4752 days to determine the
frequencies of low-degree acoustic solar oscillation modes. These
frequencies have been used to calculate the fine or small frequency
spacings and frequency separation ratios for the Sun. These spacings
and ratios have then been used to test different solar models to
investigate whether or not we can constrain solar abundances.

We find that models constructed with the older Grevesse \& Sauval
(1998) mixture satisfy the fine-spacing and separation-ratio
constraints much better than do models with the newer Asplund et
al. (2005) abundances. In fact, models with high metallicity
constructed with outdated opacities and equation-of-state data also
fare better than do the recent models with Asplund et al. (2005)
abundances.  Our investigation shows that the fine spacings and
separation ratios depend sensitively on the metallicity, and
low-metallicity models can be ruled out. 

Our results lead us to conclude that uncertainties on predicted solar
neutrinos, which had been raised because of the Asplund et al. (2005)
abundances, can now be lowered.

\acknowledgements
This paper utilizes data collected by the Birmingham
Solar-Oscillations Network (BiSON), which is funded by the UK Particle
Physics and Astronomy Research Council (PPARC). We thank the members
of the BiSON team, colleagues at our host institutes, and all others,
past and present, who have been associated with BiSON. GAV
acknowledges the support of PPARC. The radio flux observations are
made at Penticton by the National Research Council of Canada and are
available from the World Data Center. SB acknowledges partial support
from NSF grant ATM-0348837. She would also like to thank the BiSON
group for their hospitality during the time this work was carried out.
AMS is partially supported by the NSF (grant PHY-0503684), the
Association of Members of the Institute for Advanced Study 
and the W. M. Keck Foundation through a grant-in-aid
to the Institute for Advanced Study.

\clearpage

\begin{deluxetable}{llllll} 
\tablecolumns{6} 
\tablewidth{0pc} 
\tablecaption{Difference in fine-structure spacings (BiSON minus models)} 
\tablehead{ 
\colhead{}    &  \multicolumn{2}{c}{$d_{02}(n)$} &   \colhead{}   & 
\multicolumn{2}{c}{$d_{13}(n)$} \\ 
\cline{2-3} \cline{5-6} \\ 
\colhead{Model}& \colhead{Weighted mean}& \colhead{Weighted}& 
\colhead{}& \colhead{Weighted mean}& \colhead{Weighted}\\
\colhead{}& \colhead{difference ($\rm \mu Hz$)}& \colhead{RMS ($\rm \mu Hz$)}& 
\colhead{}& \colhead{difference ($\rm \mu Hz$)}& \colhead{RMS ($\rm \mu Hz$)}}
\startdata 
          JCDS& $-0.058$ ($-10.8\sigma$)& 0.072 ($13.3\sigma$)& & $-0.033$ ($ -4.5\sigma$)& 0.053 ($ 7.2\sigma$)\\ 
           SAC& $-0.133$ ($-24.7\sigma$)& 0.139 ($25.9\sigma$)& & $-0.083$ ($-11.2\sigma$)& 0.094 ($12.8\sigma$)\\ 
          BP04& $-0.009$ ($ -1.7\sigma$)& 0.047 ($ 8.7\sigma$)& & $-0.043$ ($ -5.8\sigma$)& 0.091 ($12.3\sigma$)\\ 
         BP04+& $-0.102$ ($-18.9\sigma$)& 0.126 ($23.4\sigma$)& & $-0.204$ ($-27.7\sigma$)& 0.237 ($32.2\sigma$)\\ 
      BS05(OP)& $-0.051$ ($ -9.5\sigma$)& 0.065 ($12.0\sigma$)& & $-0.083$ ($-11.3\sigma$)& 0.114 ($15.5\sigma$)\\ 
  BS05(AGS,OP)& $-0.230$ ($-42.6\sigma$)& 0.243 ($45.1\sigma$)& & $-0.360$ ($-48.7\sigma$)& 0.382 ($51.7\sigma$)\\ 
       BBS05-3& $-0.034$ ($ -6.2\sigma$)& 0.063 ($11.7\sigma$)& & $-0.078$ ($-10.5\sigma$)& 0.136 ($18.4\sigma$)\\ 
  S06+(AGS,OP)& $-0.145$ ($-27.0\sigma$)& 0.160 ($29.7\sigma$)& & $-0.200$ ($-27.1\sigma$)& 0.267 ($36.2\sigma$)\\ 
\enddata 
\label{tab:d}
\end{deluxetable} 


\begin{deluxetable}{llllll} 
\tablecolumns{6} 
\tablewidth{0pc} 
\tablecaption{Difference in frequency separation ratios (BiSON minus models)} 
\tablehead{ 
\colhead{}    &  \multicolumn{2}{c}{$r_{02}(n)$} &   \colhead{}   & 
\multicolumn{2}{c}{$r_{13}(n)$} \\ 
\cline{2-3} \cline{5-6} \\ 
\colhead{Model}& \colhead{Weighted mean}& \colhead{Weighted}& 
\colhead{}& \colhead{Weighted mean}& \colhead{Weighted}\\
\colhead{}& \colhead{difference (\%)}& \colhead{RMS (\%)}& 
\colhead{}& \colhead{difference (\%)}& \colhead{RMS (\%)}}
\startdata 
          JCDS& $-0.43$ ($ -9.5\sigma$)& 0.63 ($13.7\sigma$)& & $-0.03$ ($ -0.8\sigma$)& 0.23 ($ 5.9\sigma$)\\ 
           SAC& $-0.99$ ($-21.6\sigma$)& 1.08 ($23.5\sigma$)& & $-0.22$ ($ -5.6\sigma$)& 0.29 ($ 7.5\sigma$)\\ 
          BP04& $+0.03$ ($ +0.6\sigma$)& 0.26 ($ 5.7\sigma$)& & $+0.02$ ($ +0.4\sigma$)& 0.17 ($ 4.5\sigma$)\\ 
         BP04+& $-0.85$ ($-18.6\sigma$)& 0.92 ($20.0\sigma$)& & $-0.84$ ($-22.1\sigma$)& 0.89 ($23.2\sigma$)\\ 
      BS05(OP)& $-0.39$ ($ -8.5\sigma$)& 0.46 ($10.0\sigma$)& & $-0.29$ ($ -7.5\sigma$)& 0.35 ($ 9.2\sigma$)\\ 
  BS05(AGS,OP)& $-2.02$ ($-43.9\sigma$)& 2.06 ($44.8\sigma$)& & $-1.76$ ($-45.8\sigma$)& 1.80 ($46.9\sigma$)\\ 
       BBS05-3& $-0.05$ ($ -1.1\sigma$)& 0.27 ($ 5.9\sigma$)& & $+0.02$ ($ +0.4\sigma$)& 0.22 ($ 5.7\sigma$)\\ 
  S06+(AGS,OP)& $-0.98$ ($-21.3\sigma$)& 1.04 ($22.5\sigma$)& & $-0.58$ ($-15.1\sigma$)& 0.79 ($20.6\sigma$)\\ 
\enddata 
\label{tab:sr}
\end{deluxetable}


\clearpage

\clearpage

\clearpage

\clearpage

\clearpage

\clearpage

\clearpage

\clearpage

\end{document}